\documentclass[twocolumn]{aastex631}

\newcommand{\hi}{\mbox{H\,{\sc i}}}
\newcommand{\hii}{\mbox{H\,{\sc ii}}}

\graphicspath{{./}{figures/}}

\accepted{March 31, 2023}

\shorttitle{VLA \hi\ 21 cm Kinematics and Dynamics in NGC 5595 and NGC 5597}
\shortauthors{Garcia-Barreto \& Momjian (2023)}

\begin{document}
\title{\hi\ 21 cm Extended Structures to the North-East, and South-West of NGC 5595: VLA Observations of the Disk Galaxy Pair NGC 5595 and NGC 5597}

\correspondingauthor{J. Antonio Garcia-Barreto}
\email{j.antonio.garcia.barreto@astro.unam.mx}

\author[0000-0002-3773-9613]{J. Antonio Garcia-Barreto}
\affil{Instituto de Astronomia,
Universidad Nacional Aut\'onoma de M\'exico, \\
Apartado Postal 70-264, Ciudad de M\'exico 04150, M\'exico}

\author[0000-0002-3168-5922]{Emmanuel Momjian}
\affil{National Radio Astronomy Observatory,
P. V. D. Science Operations Center, \\
P.O. Box O, Lopezville Road, Socorro, 87801-0387, New Mexico, U.S.A.}

\begin{abstract}
We report VLA B-configuration observations of the \hi\ 21 cm line on the close disk galaxy pair NGC 5595 and NGC 5597. At the angular resolution of the observations, $\sim7\farcs1 \times 4\farcs2$, while most of the \hi\ 21 cm in NGC 5595 and in NGC 5597 has the same extent as the optical disk, we have detected for the first time extended structures (streamers) to the north-east (NE), and south-west (SW) of NGC 5595 with no counterparts in blue, red optical (continuum), 20 cm radio continuum, or H$\alpha$ spectral-line emission. One structure is extended by $\sim 45''$ to the NE with blue-shifted velocities, and the other by $\sim 20''$ to the SW with red-shifted velocities with respect to the systemic velocity. No \hi\ 21 cm emission is detected from the innermost central (nuclear) regions of either galaxy. Lower angular resolution \hi\ 21 cm imaging indicates the non-existence of any intergalactic \hi\ 21 cm gas as tails or bridges between the two galaxies. Our new 20 cm radio continuum emission image of NGC 5597 shows a strong unresolved elongated structure at the central region, in the north-east south-west direction, very similar to the spatial location of the innermost H$\alpha$ spectral line emission. There is no 20 cm continuum emission from its north spiral arm. In NGC 5595, the 20 cm radio continuum image shows no continuum emission from the NE nor the SW extended structures with \hi\ 21 cm emission. 
\end{abstract}

\keywords{galaxies: active: individual (NGC 5595, NGC 5597) ---galaxies: kinematics and dynamics --- galaxies: ISM --- galaxies: spiral}

\section{Introduction}  \label{sec:intro}

From the observational point of view, ongoing major and minor gravitational interactions of two or more disk galaxies show extended \hi\ 21 cm cold gas structures of tidal origin, with M51 and NGC 5195 being one of the classic examples \citep{san99}.

The numerical simulations performed by citet{too72} have convincingly shown that the gravitational interactions in very close pair of disk galaxies in their late stages of merging cause stellar tails and bridges (e.g., M51 with NGC 5195, NGC 4038 with NGC 4039, NGC 4676A with NGC 4676B -- the mice). They went even further and dared to raise the possibility that the innermost density-wave spiral pattern in galaxies such as M51 and NGC 7753 was caused indirectly by the recent external influences (see their section VII subsection {\it d)} ending with the question, {\it Et tu} M81? referring to M81 with the close companion NGC 3077). 

Since then gravitational interaction between a close pair of disk galaxies has been recognized to be important in determining their gas kinematics, dynamics, star formation, central feeding of a massive black hole and their evolution in time \citep{mih94,mih96}. 

Recent galaxy distribution surveys have confirmed that galaxies are found in a hierarchical structure of filaments and walls that surround large galaxy voids with 54\% in the local universe concentrated in virialized clusters and groups, and only 7\% as isolated pairs of galaxies \citep{arg15}. Galaxy pairs detected in the far infrared by the Infrared Astronomical Satellite (IRAS) have a median separation of 20\,$h^{-1}$\,kpc, and pairs that are most probably close together in space have greater specific star formation rates (SSFR) \citep{gel06}. Using H$\alpha$ line emission to estimate star formation rate in major mergers\footnote{Major mergers have been defined as pair of galaxies whose difference in apparent magnitude is less than 2, $\Delta m \leq 2$; minor mergers on the contrary are pair of galaxies whose difference in apparent magnitude is more than 2, $\Delta m \geq 2$ \citep{woo07}.} from 2409 galaxies reveals enhanced star formation in cases where the gravitational tidal force is relatively strong compared to the self-gravity of a galaxy \citep{woo07}. These authors also found that for major mergers there is a correlation between SSFR and separation (on the plane of the sky), $\Delta D$, in such a way that the smaller $\Delta D$ the higher the SSFR \citep{woo07}.

Radio continuum observations at $\lambda \sim 11$ cm, $\lambda \sim 6$ cm and $\lambda \sim 3.7$ cm of disk-disk pairs from the Catalog of Isolated Pairs of Galaxies (CPG; \citealt{kar72}) revealed a correlation between physical separation and radio emission, with close pairs being radio emitting sources at more than twice the number than more widely spaced pairs \citep{sto78a,sto78b}. Westerbork radio interferometer continuum observations at 1.415\,GHz have shown that the radio power of the central sources in pair of galaxies is on average four times higher than in isolated disk galaxies \citep{hum81,hum87}. Furthermore, Very Large Array radio continuum observation at 1.46\,GHz and 4.88\,GHz of 60 {\it interacting} galaxies have shown that on average the radio power of the central radio sources are about a factor of five higher than in isolated galaxies, and that the radio power of central radio sources in barred disk galaxies are a factor of five higher than in non-barred disk galaxies \citep{hum87,hum90}. 

What \citet{too72} did not explicitly state was that narrow and long tails and bridges of cold \hi\ 21\,cm gas are also caused by gravitational interaction of close pairs of disk galaxies. Indeed, three of the most beautiful examples of \hi\ 21 cm imaging of interactions in a system of galaxies are: i) NGC 5194 (M51, Sbc(s)I-II) with its SE long tail and its companion NGC 5195 (SB01 pec) \citep{dav74,hay78,app86,rot90}, ii) the M81 (Sb(r)I -II) -- M82 (amorphous) and NGC 3077 (Irr) system with cold atomic hydrogen gas structures from a northern tidal bridge between M82 and NGC 3077, a southern tidal bridge between NGC 3077 and M81, and the structure between M81 and M82 \citep{cot76,van79,yun93,yun94}, and iii) NGC 4038/4039 ``the antennae'', NGC 4038 SB,(s)mpec NGC 4039 SAB(s)mpec \citep{van79,gor01,hib01}. 

One such nearby disk galaxy pair system is NGC 5595 and NGC 5597. For this system, we adopt a Hubble (spectroscopic) distance of $D_{{\rm pair}} = 38.6$\,Mpc \citep{tul88}, giving an approximate angular-to-linear scale of $1\farcs0 \sim 187.14$ pc. The projected separation (on the plane of the sky) between the two galaxies is $\sim 3\rlap{.}{'}97$, or $\sim 45$\,kpc.  Our analysis of the \hi\ 21 cm gas emission, rotation curve, and angular velocity in NGC 5597 using data obtained with the Karl G. Jansky Very Large Array (VLA)\footnote{The National Radio Astronomy Observatory is a facility of the National Science Foundation operated under cooperative agreement by Associated Universities, Inc.} in B-configuration, and the subsequent estimation of the angular velocity pattern of its stellar bar have already been reported \citep{gar22}.
   
Since NGC 5595 and NGC 5597 are so close on the plane of the sky, it required a single VLA pointing (primary beam full width at half power, $\theta_{\rm HPBW} = 30'$ at 1.4\,GHz) to observe the \hi\ 21 cm emission from the disk galaxy pair system. Therefore, here we report the overall \hi\ 21 cm kinematics and dynamics of each disk galaxy and their close environment.

NGC 5597 was originally chosen from a list of 56 nearby, bright, barred galaxies with IRAS 60$\mu$m flux densities greater than 5 Jy \citep{gar96} and colors indicative of circumnuclear star formation \citep{hel85}. Its IRAS flux densities are $f_{60\mu{\rm m}} \sim 8.7$ Jy, $f_{100\mu{\rm m}} \sim 15.32$ Jy with $T_{dust} \sim 36\degr$ K. In 1996, we were only aware that NGC 5597 was part of Tully's group 41+15+15. Later, we carried out the study of companions for the same set of nearby bright barred galaxies utilizing NED\footnote{The NASA/IPAC Extragalactic Database (NED) is operated by the Jet Propulsion Laboratory, California Institute of Technology, under contract with the National Aeronautics and Space Administration.} database and found out that NGC 5597 has only one nearby disk galaxy companion, namely NGC 5595 \citep{gar03}. Based on their projected spatial proximity ($\Delta D_{12}/(d_1 + d_2) \leq 2$   \citep{zwi56,kar72,kar81}) and very similar spectroscopic \hi\ 21 cm systemic velocities ($\Delta V \leq 5$ km s$^{-1}$), NGC 5597 forms an isolated gravitationally bound pair with the disk galaxy NGC 5595 (Sc). 
    
There is also growing evidence of weak nuclear activity in normal disk galaxies with and without a prominent stellar bar, with observational detection of nuclear low velocity bipolar outflows in H$\alpha$, e.g., M81 \citep{goa76}, NGC 1068 \citep{ulv87}, M51 \citep{for85,cec88,cra92,sco98}, NGC 3079 \citep{vei94}, M101 \citep{moo95}, NGC 3367 \citep{gar98,gar02}, and NGC 1415 \citep{gar19}, suggesting that they represent a low end of scale for nuclear activity after quasars, BL Lac objects, radio galaxies, and Seyfert galaxies. All such phenomena have an origin in events occurring around central massive black holes with the level of the activity being governed by the gas supply to fuel these central engines \citep{nor83}. Similar physical processes have also been observed in NGC 3367 \citep{gar98,gar02,gar05,her11}).

In this paper, we present VLA B-configuration observations of atomic hydrogen, \hi\ 21 cm, cold gas emission at an angular resolution of $7\farcs1 \times 4\farcs2$ at P.A.$\sim -10\degr$ EofN (or about $1.33 \times 0.78$ kpc in linear size), as well as new 20 cm radio continuum emission from the disk galaxy pair NGC 5595 and NGC 5597. We have imaged the global \hi\ 21 cm emission from this galaxy pair to probe the internal kinematics, and investigate the existence of any large scale gas structure between them. We have also made 20 cm radio continuum emission images for both disk galaxies. For the barred galaxy NGC 5597, we compare its 20 cm continuum emission with previously published H$\alpha$ spectral line emission and blue optical (broad band continuum from 103aO glass plate) spatial distribution, especially in the inner 30$\farcs0$. This work is organized as follow: $\S$ 2 presents the observations and data reduction. $\S$ 3 presents the disk galaxy system as an isolated gravitationally bound pair. $\S$ 4 presents some properties of NGC 5597 as an SBc disk galaxy. $\S$ 5  presents some properties of NGC 5595 as an Sc disk galaxy, $\S$ 6 presents the \hi\ 21 cm features in the field of the disk galaxy pair NGC 5595 and NGC 5597, and finally $\S$ 7 presents the summary and conclusions. 

\section{Observations and Data Reduction} \label{sec:observations}

    Two dimensional velocity fields for the \hi\ 21 cm emission from the
disk galaxy pair NGC 5595 and NGC 5597 were obtained using the VLA in its B configuration on 2019 June 6, 7, 11, 16, 18, \& 22. In each day, the total observing time was about $1.5^h$ and included 100 min of on-source time, as well as overhead to observe the flux density scale calibrator/bandpass calibrator 3C\,286, and the complex gain calibration J1448$-$1620. The VLA system was tuned to the rest frequency of the \hi\ 21 cm line $\nu_{\rm rest} = 1,420,405.752$\,kHz redshifted to a mean heliocentric velocity of $v_{\rm pair} = 2,700$\,km s$^{-1}$ for the galaxy pair system. Table 1 lists the coordinates, Hubble type, distance, and systemic \hi\ velocities of the two galaxies NGC 5595 and NGC 5597. Furthermore, because these two galaxies are separated by only $\leq 4'$ on the plane of the sky, and are well within the primary beam of the VLA antennas ($\theta_{\rm HPBW} = 30'$ at 1.4\,GHz), the pointing center of the observations was chosen to be midway between the two disk galaxies: $\alpha_{mid} = 14^{\rm h}\,24^{\rm m}\,21\rlap{.}{^{\rm s}}0, \delta_{mid} = -16\degr\,44'\,45\farcs0$.

\startlongtable
\begin{deluxetable*}{lllcllcccc}
\tabletypesize{\normalsize}
\tablecaption{NGC 5595 - NGC 5597 Pair of Disk Galaxies: Coordinates, Hubble type, Distance, Systemic \hi\ Velocity}  \label{tab:table1} 
\tablehead{
\colhead{Galaxy} & \colhead{$\alpha$(J2000)} & \colhead{$\delta$(J2000)} & \colhead{Ref.} & \colhead{RSA type} & \colhead{NED type} & \colhead{Distance} & \colhead{Ref.} & \colhead{$V(\hi)_{sys}$} & \colhead{Ref.} \\
\colhead{Name}  & \colhead{$hh~mm~ss$}    & \colhead{$\degr~~' ~~\farcs$} & \colhead{} & \colhead{} & \colhead{} & \colhead{Mpc}  & \colhead{} & \colhead {km s$^{-1}$} & \colhead{} }
\colnumbers
\startdata
NGC 5595 &  $14~24~13.3$ & $-16~43~21.6$ & 1 & Sc(s) II & SAB(rs)c & 38.6 & 2 & 2697 & 3\\
NGC 5597 &  $14~24~27.49$ & $-16~45~45.9$ & 1 & SBc(s) II & SAB(rs)b & 38.6 & 2 & 2698 & 3\\
\hline 
\enddata
\tablecomments{1) \citep{dia09}, 2) \citep{tul88}, 3) \citep{pat03} }
\end{deluxetable*}

The observations utilized one of the 1\,GHz wide 8-bit sampler pairs of the VLA, namely A0/C0, and the Wideband Interferometric Digital ARchitecture (WIDAR) correlator was configured to deliver a single 4\,MHz wide subband with 256 spectral channels, resulting in a channel spacing of 15.625\,kHz or 3.3\,km s$^{-1}$. This subband spans a velocity range of $\Delta V \sim 843$ km s$^{-1}$, which is sufficient to cover the full width at zero intensity of the \hi\ emission from both NGC 5595 and NGC 5597, and provide line-free channels for continuum subtraction.

    We used the Common Astronomy Software Applications (CASA) for the
flux density scale, bandpass, complex gain calibration, continuum subtraction, as well as deconvolution and imaging. The Astronomical Image Processing System (AIPS) was utilized for the spectral and kinematics analysis, and for the presentation and overlay of 20 cm continuum, H$\alpha$ and optical 103aO images.
    
The data set of each session was calibrated independently, and image cubes were made to determine the line-free channels for continuum subtraction. Line+continuum emission from NGC 5597 was confined between channels 81 to 200, and from NGC 5595 between channels 81 and 197. The continuum emission was subtracted from the line+continuum data in the $uv$ plane. The \hi\ image cube was then produced by combining the data from all the sessions, using a cell size of 1$\farcs0$ and Briggs weighting with robust=0.8. The resulting synthesized beam was $\sim 7\farcs1 \times 4\farcs2$ ($1.33 \times 0.78$ kpc$^2$) at full width half maximum (FWHM) with P.A.$\sim -10\degr$ EofN. This image cube was imported to AIPS for further analysis. In there, Moments 0, 1, and 2 were made using a flux density cutoff of 2.5$\sigma$, with 1$\sigma \sim 450\,\mu$Jy beam$^{-1}$ channel$^{-1}$. The \hi-line-free 20 cm continuum emission, using the channel ranges 45--75 and 222--240 from all the sessions, was also imaged in CASA using robust=0. The resulting synthesized beam (FWHM) was $\sim 6\farcs1 \times 3\farcs7$ at P.A.$\sim -8\degr$ EofN, and the rms noise was 1$\sigma \sim 167.4\,\mu$Jy beam$^{-1}$.

    To search for extended \hi\ 21 cm cold atomic gas in between NGC
5595 and NGC 5597, we have also made a low resolution image cube using a $u, v$ range restricted between 0 and 5\,k$\lambda$. This resulted in a synthesized beam (FWHM) of $ \sim 30\farcs7 \times 28\farcs8$ at P.A. $\sim +50\degr$ EofN, corresponding to a linear resolution of $\sim 5.4 \times 5.7$ kpc$^2$.

\section{Gravitationally bound isolated pair: NGC 5597 - NGC 5595}

    The disk galaxies NGC 5595 and NGC 5597 are in the Southern Virgo -
Libra Cloud of galaxies, Tully's group 41-14(+14), with a very low galaxy volume density of only 0.16 galaxies Mpc$^{-3}$; see their spatial location at galactic coordinates $l^{II} \sim 332\degr.8$, $b^{II} \sim +40.7$ in Plates 1, and 5 \citep{tul87}, and in $\alpha$, and $\delta$ in Fig. 2 \citep{tam85}. 

    These two galaxies are separated by $\Delta\alpha \sim
13\rlap{.}{^s}54, \Delta\delta \sim 138''$ or $D_{12} \sim 3\rlap{.}{'}97$ on the plane of the sky. The diameters of NGC 5597 and NGC 5595 are $d_1 \sim 1\rlap{.}{'}87$ and $d_2 \sim 1\rlap{.}{'}6$, respectively. Furthermore, these two galaxies have very similar spectroscopic \hi\ 21 cm systemic velocities ($\Delta V < 5$ km s$^{-1}$). Thus, they indeed satisfy Holmberg's spatial and velocity difference criteria $\Delta D_{12}/(d_1 + d_2) \ll 2$ \citep{zwi56,kar72,kar81} indicating that NGC 5597 forms an isolated gravitationally bound pair with the disk galaxy NGC 5595 (Sc).

\begin{figure*}[htb!]
\includegraphics[width=15cm,height=14cm]{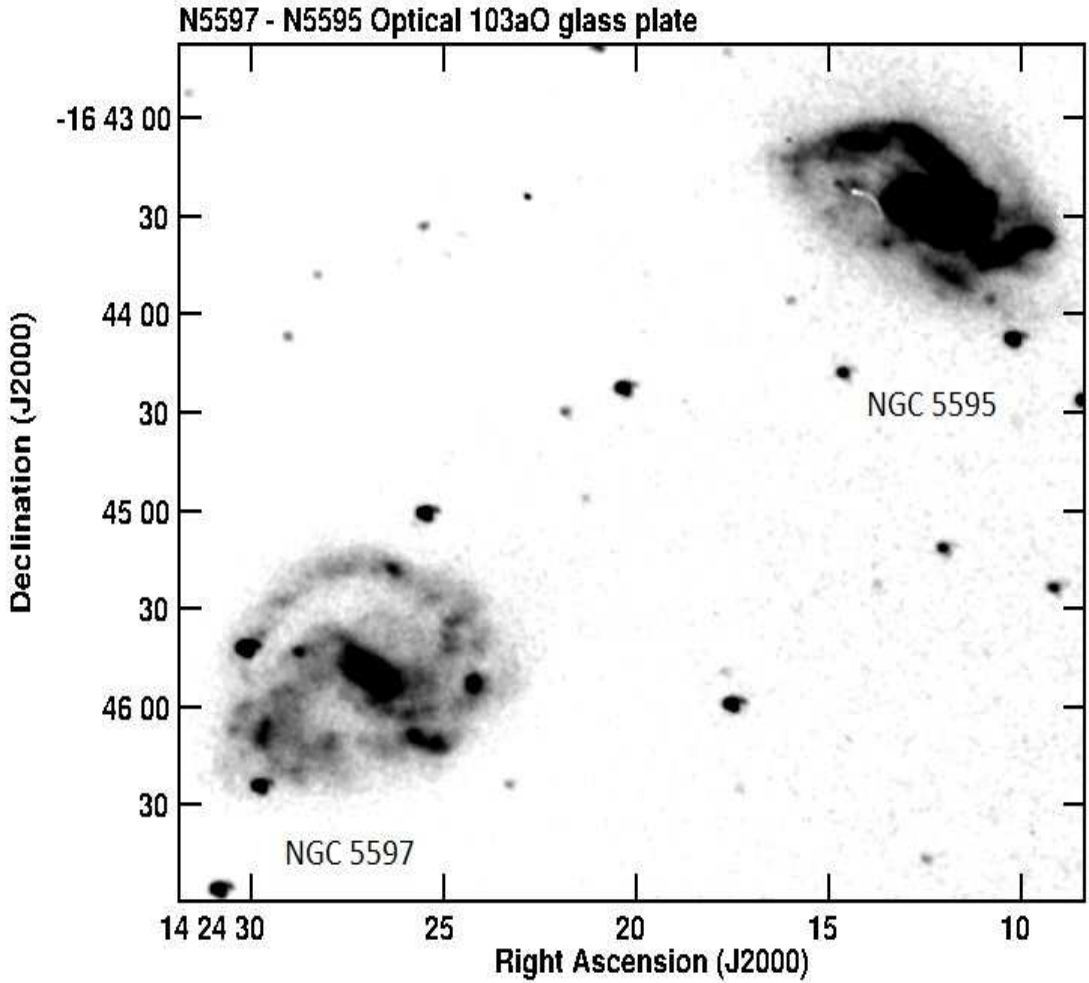}
\figcaption{Reproduction of optical blue continuum image (glass plate 103aO) from the close pair of disk galaxies NGC 5597 at the SE and NGC 5595 at the NW, obtained with the OAN-SPM 2.1m optical telescope in Mexico \citep{dia09}. The flux density scale has not been calibrated. Grey scale is from $2 \rightarrow 27.8 \sigma$, where $\sigma$ is the noise in arbitrary units.  \label{fig. 1}}
\end{figure*}

    Figure 1 shows a reproduction of the optical blue continuum 103aO
emission from the disk galaxy pair NGC 5597 (SE) and NGC 5595 (NW) in grey scale in relative units taken with the OAN-SPM 2.1m optical telescope in Mexico \citep{dia09}. In this blue optical continuum image there are no bridges or structures connecting the two galaxies. In Section 5, it will be shown that there are \hi\ 21 cm extended cold gas structures to the NE and SW of NGC 5595. 

\section{NGC 5597: a late type barred disk galaxy}

    NGC 5597 is a bright (m$_B \sim 12.57$) disk galaxy classified as
SBc(s) II \citep{san81}, and as SAB(s)cd, in NED. Table 2 lists the spatial positions of its nucleus estimated from observations at different wavelengths and Table 4 gives its basic properties.

\begin{figure*}[bht!]
\begin{center}
\includegraphics[width=12cm,height=12cm]{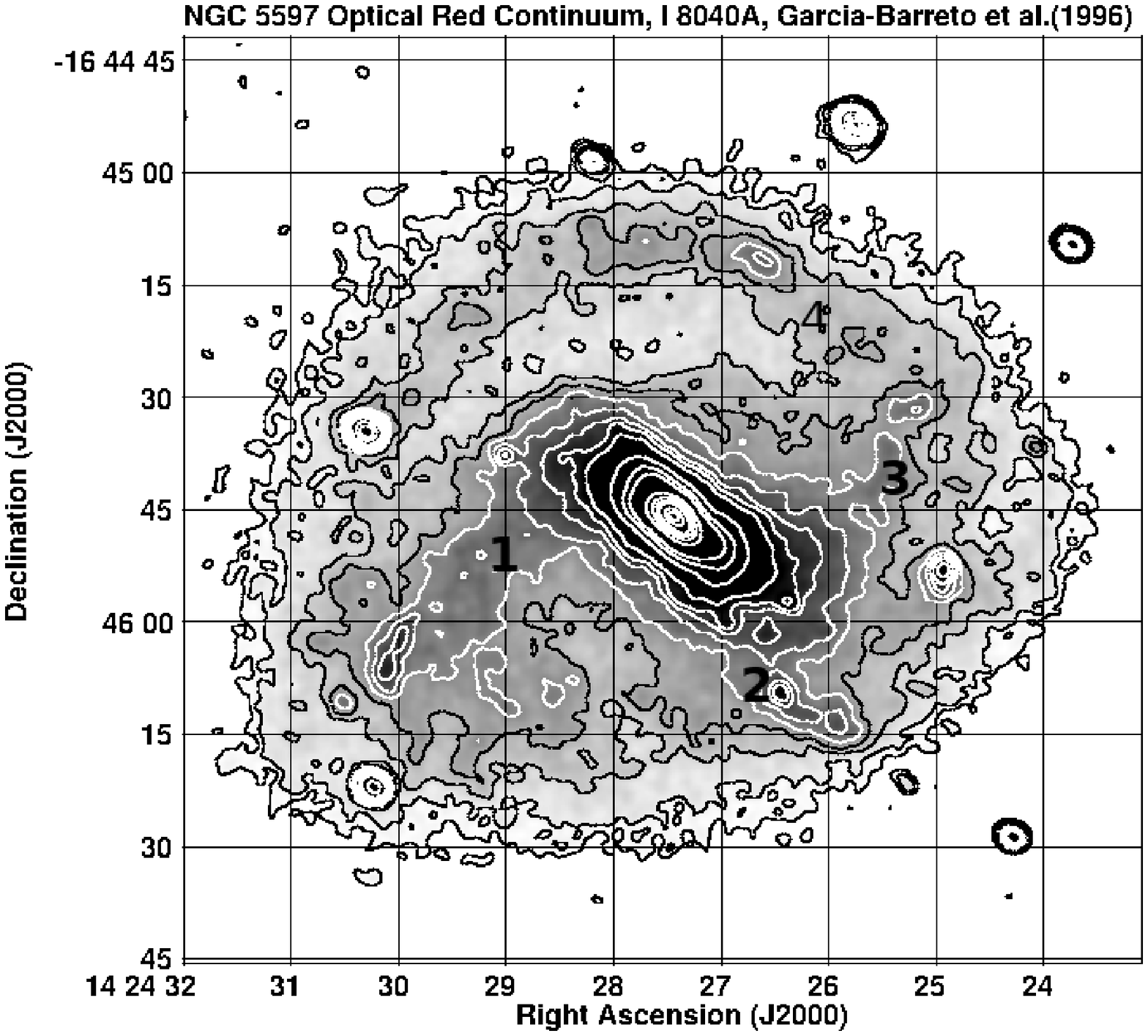}
\figcaption{Reproduction of optical red continuum image (broadband filter I 8040 \AA), in grey scale and contours, of the disk galaxy NGC 5597 obtained with the OAN-SPM 2.1m optical telescope in Mexico \citep{gar96}. The image has been convolved with a circular Gaussian beam of $\sim 1\farcs5$ at FWHM. The flux density scale has not been calibrated, thus the grey scale stretch is from $5 \rightarrow 37 \sigma$, where 1$\sigma \sim 45$ is the noise in arbitrary units. Contours are at 5, 7, 11, 15, 20, 25, 30, 34, 40, 60, 80, 100, 200, 300, 400, 500, 750, 900 $\times 1\sigma$. \label{fig. 2}}
\end{center}
\end{figure*}

\begin{figure*}[bht]
\includegraphics[width=6cm,height=6cm]{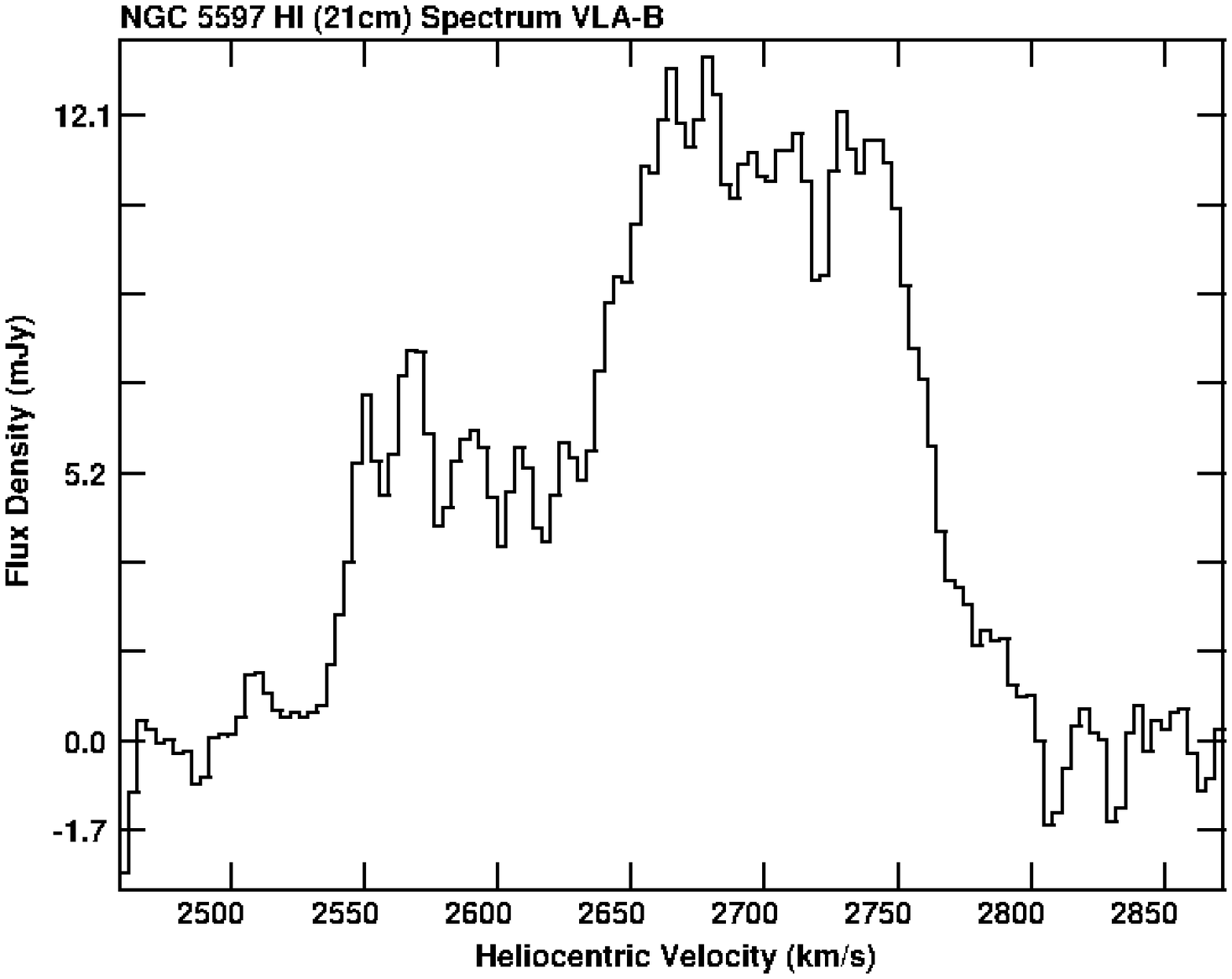}
\includegraphics[width=6cm,height=6cm]{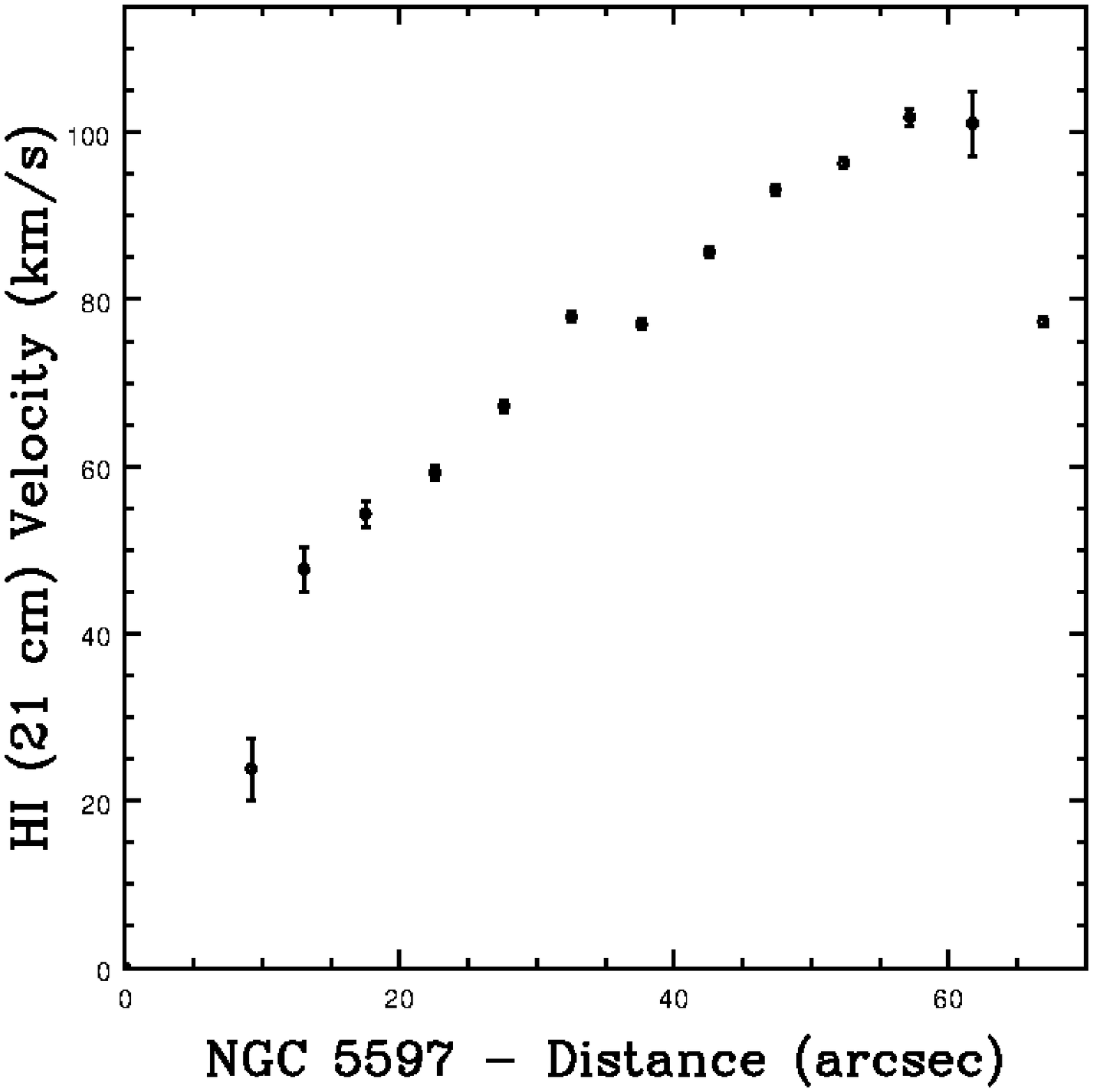}
\includegraphics[width=6cm,height=6cm]{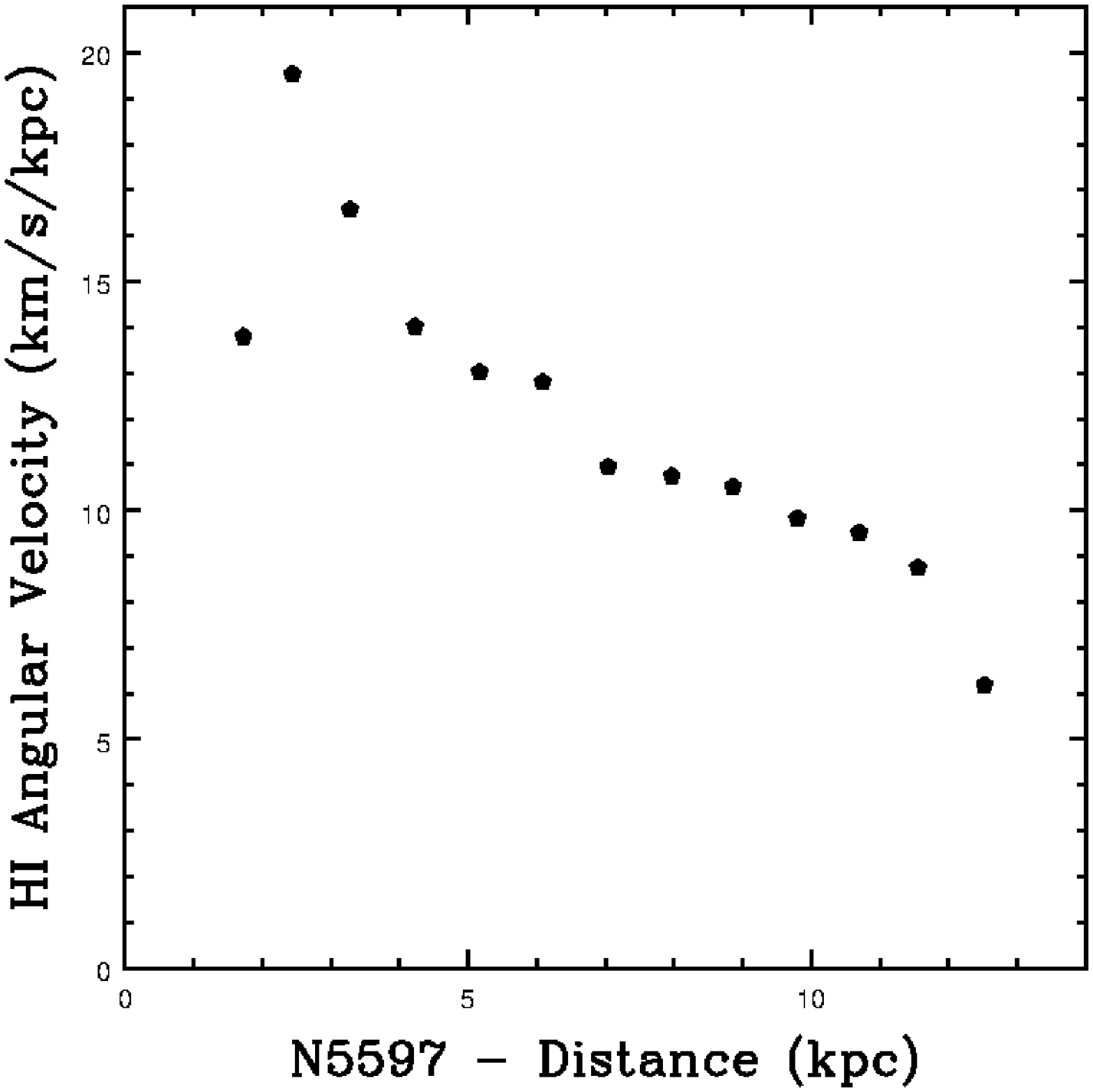}
\figcaption{Left panel: \hi\ 21 cm spectrum of NGC 5597 obtained from VLA B-configuration observations. The heliocentric systemic velocity, fitted by the task {\it GAL} in AIPS, is $V({\rm sys})_{\rm helio} = 2698$ km s$^{-1}$, with $\Delta V_{50\%} \sim 211.40$ km s$^{-1}$, and $\Delta V_{20\%} \sim 239.25$ km s$^{-1}$. The flux density scale is very difficult to compare to previous Parkes (64 m) and Green Bank (91m) single-dish spectra, because their beam included both disk galaxies NGC 5595 and NGC 5597 \citep{mat92,spr05}. The shape of the spectrum looks very similar to that obtained with Nan{\c{c}}ay, however its beam at FWHM was $\sim 3\rlap{.}{'}6$ east-west $\times~22'$ north-south \citep{pat03}. Central panel: \hi\ 21 cm rotation curve of NGC 5597. Right panel: \hi\ 21 cm angular velocity curve of NGC 5597.
\label{fig. 3}}
\end{figure*}

    Figure 2 shows a reproduction of the red optical (filter I 8040 
\AA) of the disk galaxy NGC 5597\footnote{The red optical image has not been calibrated in flux density scale, therefore the contours are proportional to the rms noise in relative units.}.
Both images of NGC 5597 in Figures 1 and 2 show a central elongated structure about $2a \sim 28\farcs3$ by $2b \sim 14''$ at a P.A. $\sim 52\degr$ that we interpret as the boxy stellar bar and have estimated its angular velocity pattern ($\Omega_{\rm bar}$) \citep{gar22}. 

    Additionally, as seen in the SE galaxy in Figure 1, and more
specifically in Figure 2, there are at least four narrow and curved structures as spiral arms outside the central region of NGC 5597 (labeled 1 to 4 in Figure 2). The first structure (1) starts from the NE half of the stellar bar and extends to the SE, the second structure (2) starts from the SW of the southern half of the stellar bar and extends further to the SW, the third structure (3) starts from the NW of the southern half of the boxy bar and extends further NW joining the outer fourth structure. 

    The fourth structure (4) seems to start from the SW of the disk and
continues counter clockwise to N, NE, and SE just parallel on the outside of the first structure. A plausible interpretation of the north optical spiral arm in NGC 5597 would be that it is an spiral arm near an outer ultraharmonic resonance, UHM\footnote{For a star (gas cloud) in a axisymmetric gravitational potential of a disk galaxy, $\Phi_{\rm disk}$, in addition of a weak non-axisymmetric gravitational potential, $\Phi_{\rm bar}$, where $\Phi_{\rm bar} << \Phi_{\rm disk}$, its orbit can be represented as a superposition of the circular motion of a guiding center and small radial oscillations around this guiding center. In cylindrical coordinates with z=0, the stellar bar gravitational potential may be expressed as $\Phi_{\rm bar}(R,\varphi,z=0) = \Phi_{\rm b} cos(m\varphi)$. The equation of motion of a star (gas cloud) is given by the equation of motion of a harmonic oscillator of natural frequency $\kappa(R)$ that is driven at a frequency $m(\Omega_{\rm star,gas cloud} - \Omega_{\rm bar})$. The solution of the radial motion becomes singular at different values of the guiding center. At corotation radius $\Omega_{\rm star, gas cloud} = \Omega_{\rm bar}$. At other radii $m(\Omega_{\rm star, gas cloud} - \Omega_{\rm bar}) = \pm \kappa$; for $m=2$ the plus and minus signs denote the so called Inner and Outer Lindblad Resonances. When $m=4$, or $m=6$ they denote the radii of the so called ultra harmonic resonances (UHM); in particular when $m=4$ the minus sign denotes the outer UHM. \citep{bin87,con88,con89,ath92}} m=4 \citep{con89}.

    Figure 3-Left shows our VLA B-configuration \hi\ 21 cm spectrum for
NGC 5597. Its overall shape is similar only to the previous Nan\c{c}ay spectrum, because it has the approximate east - west angular resolution to isolate the emission from NGC 5597 \citep{pat03}. The full width of the \hi\ emission line at $20\%$ of the peak, seen in our VLA B-configuration spectrum of NGC 5597, is $\Delta V_{20\%} \sim 239.25$ km s$^{-1}$ while the full width at $50\%$ of the peak is $\Delta V_{50\%} \sim 211.40$ km s$^{-1}$ (see Figure 3-left). The Parkes and Green Bank single dish radio telescopes did not have enough angular resolution to separate the two galaxies, resulting in spectra that showed the combined \hi\ 21 cm emission from NGC 5595 and NGC 5597 \citep{mat92,spr05}. Figure 3-Middle shows the rotation curve of NGC 5597, after several iterations with AIPS task {\it GAL}, assuming that gas orbits are circular. This rotation curve was obtained in confocal circular anuli each of them 8$\farcs0$ wide integrated from R=$0\farcs0$ up to R$=70\farcs0$ with both hemispheres (red and blue shifted velocities compared to the systemic velocity) \citep{rog74}. It shows a slowly rising curve with a low velocity value of V$ \sim 24$ km s$^{-1}$ at $R \sim 9\farcs23$ ($\sim 1.73$ kpc) up to $V \sim 101$ km s$^{-1}$ at $R \sim 61\farcs8$. The angular velocities curve, $\Omega_{\rm gas}$, shown in Figure 3-Right decreases in such a way that it is less pronounced than a simple $\Omega_{\rm gas} \propto 1/(R^{3/2})$, perhaps suggesting an extended central mass distribution\footnote{Modeling a detailed mass distribution in NGC 5597 is beyond the scope of the present study.}.

\begin{figure*}[bht]
\includegraphics[width=8cm,height=8cm]{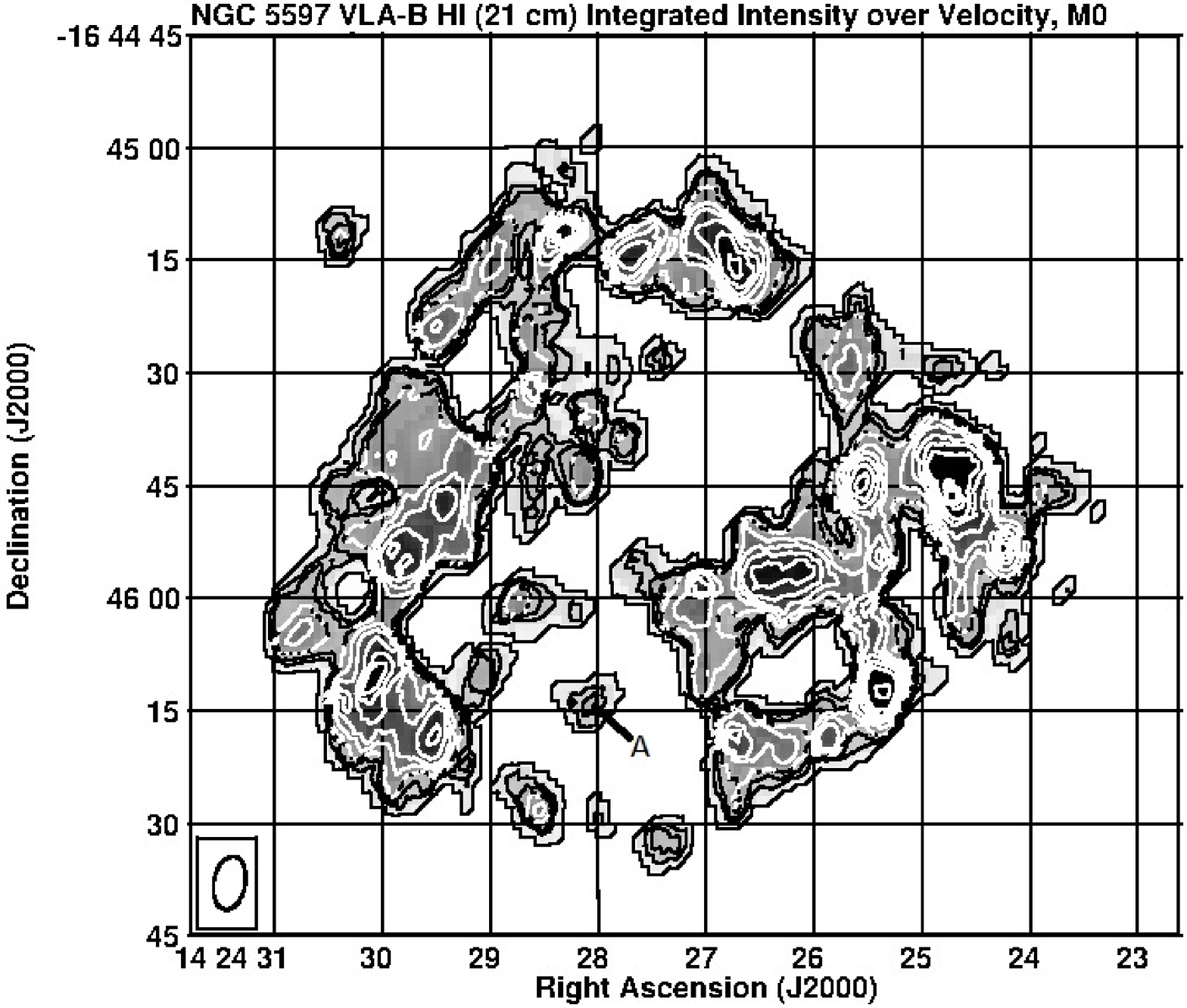}
\includegraphics[width=8cm,height=8cm]{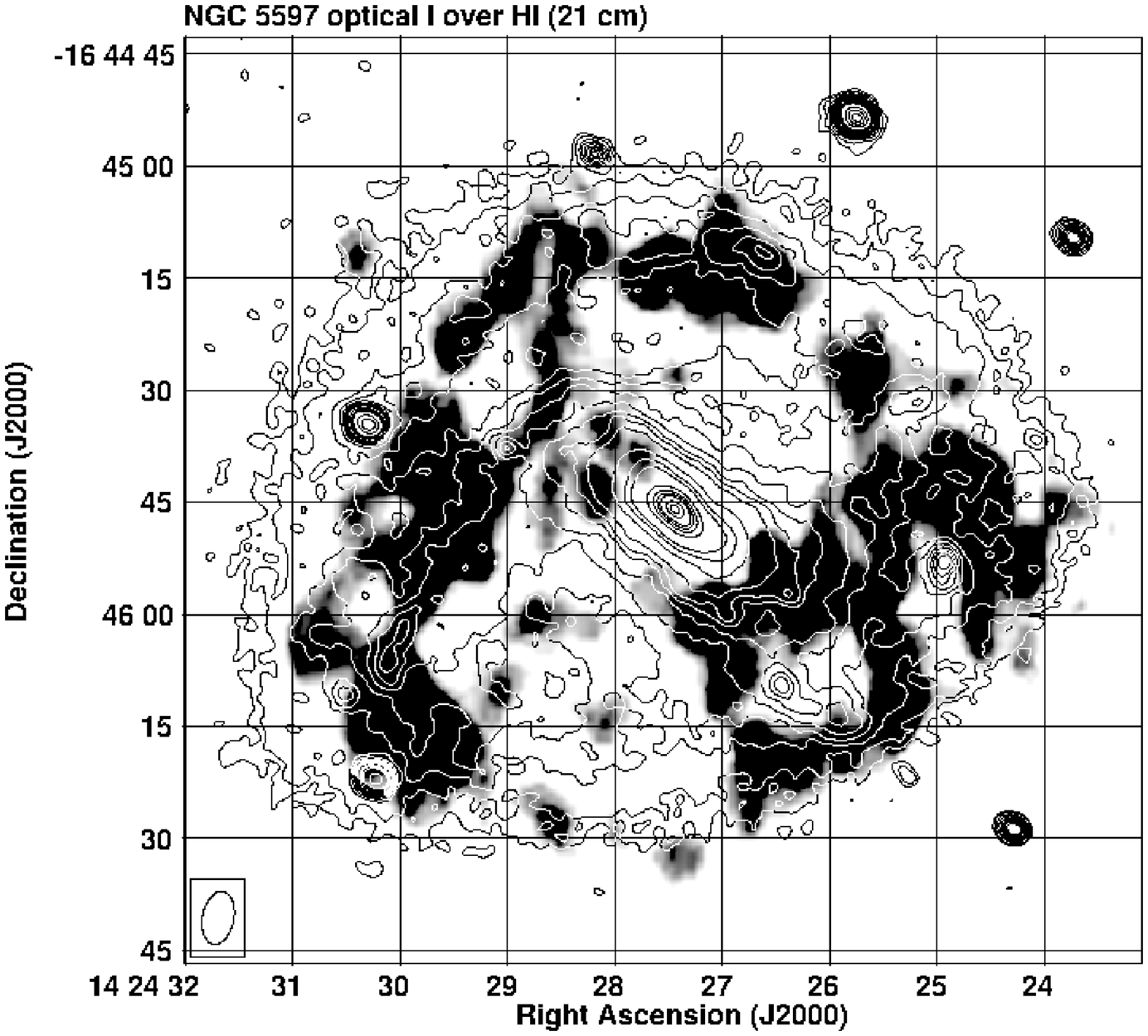}
\figcaption{Left: \hi\ 21 cm VLA B-configuration integrated intensity over velocity, Moment 0, image of NGC 5597 in contours superposed on grey scale. The contours are at 3, 35, 50, 70, 100, 125,150,175, 200, 250, 270 times $ 1 \sigma \sim 0.3$ mJy/beam  km/s. Greyscale stretch is from 3$\sigma$ to 200$\sigma$. The synthesised beam at FWHM is shown in lower left corner. Letter $A$ indicates an isolated source as if it were in the center of a circular area devoid of \hi\ emission . Right: Optical I filter image in contours (same as Figure 2) over \hi\ (21 cm) M0 in grey scale, with greyscale stretch from 2 to 17.5 mJy\,beam$^{-1}$\,km\,s$^{-1}$.
\label{fig. 4}}
\end{figure*}

\subsection{NGC 5597 VLA B-Configuration: Neutral Atomic Gas Spatial Distribution}

    At the angular resolution of our VLA observations (FWHM of $\sim
7\farcs1 \times 4\farcs2$), all of the \hi\ 21 cm emission is from within the optical disk of NGC 5597. 

    Figure 4a shows the \hi\ 21 cm VLA B-configuration integrated
intensity over velocity (Moment 0) superposed on itself in grey scale. Figure 4b shows optical I filter image in contours superposed on \hi\ ( 21 cm) M0 in grey scale. We note that all the \hi\ 21 cm cold gas emission arises from within the optical disk of the galaxy. Notice the narrow, curved and long structures in \hi\ in the NE with an abrupt decrease in emission towards the eastern side. The bright \hi\ cold gas emission from the north lies at a mean distance of $5.8 \pm 1$ kpc while the optical spiral arm (labeled 4 in Figure 2) lies at a mean distance of $6.9 \pm 0.94$ kpc on the plane of the sky. Also, there is a weak unresolved peak of \hi\ emission, labeled $A$ in Figure 4a, as if it were the center of a circular area of radius $\sim 15''$ devoid of \hi\ emission. At that spatial position there is very weak (almost no) optical red emission (see Figure 2). As a comparison, other disk galaxies generally show \hi\ 21 cm gas emission well beyond the optical disk, e.g., M83 \citep{rog74}, NGC 1300, SB(rs)bc \citep{eng89}, NGC 3783, SBa, \citep{gar99}, NGC 3147 S(rs)bc \citep{haa08}. 

    Remarkably in NGC 5597 there is no \hi\ emission from the central
region that includes the nucleus and the circumnuclear area. Another area that lacks \hi\ emission is located to the SW of the optical stellar bar and coincides with the odd SW optical spiral arm labeled 2 in Figure 2. It is worth noting that in other disk galaxies (e.g., M31) \hi\ holes are associated with \hii\ regions \citep{com02}, with hot gas of T$ \sim 10^4$ K \citep{spi78}, or regions with newly formed stars. 

    Furthermore, there is no \hi\ cold gas emission from the position of
the optical nucleus, $\alpha$(J2000) = $14^{\rm h}24^{\rm m}27\rlap{.}^{\rm s}49$, $\delta$(J2000) =$ -16\degr45'45\farcs9$, nor from the innermost circumnuclear region where the gas is hot and ionized as indicated by the H$\alpha$ continuum-free \citep{gar96} and 20 cm radio continuum emission from that region \citep{con90} and our new VLA B-configuration 20 cm continuum image (see following sections).

\subsection{NGC 5597 VLA B-Configuration: Neutral Atomic Gas Velocity Field and Kinematics}

    Our kinematical analysis of the \hi\ 21 cm emission from NGC 5597
with the task {\it GAL} in AIPS is presented in Table 5.

    Figure 5 shows the velocity field with small redshifted velocities
(Left) compared to the systemic velocity of NGC 5597 
and with large redshifted velocities (Right).
The P.A. of the redshifted semimajor axis is $\sim 100\degr$ EofN, with the corresponding P.A. of zero velocities compared to systemic at $\sim 10\degr$ EofN. 

    Figure 6-Left shows the velocity contour lines of the small
blueshifted velocities while Figure 6-Right shows the large blue shifted velocity contour lines. To first order, the velocity field indicates a normal disk\footnote{In a normal disk galaxy in the approximation with only circular orbits  ($v_R = 0$, $v_z=0$) the velocity field will be symmetric about the minor axis and one side will show redshifted velocities compared to the galaxy's systemic velocity, while the other side will show blueshifted velocities; see Fig. 8-17 in \citep{mih81} and Fig. 3.6 in \citep{com02}.} in differential rotation, with the south-west, SW, $\rightarrow$ south $\rightarrow$ south-east, SE, $\rightarrow$ north-east, NE hemisphere showing redshifted velocities, while the north-east $\rightarrow$ north $\rightarrow$ north-west $\rightarrow$ west hemisphere showing blueshifted velocities.

    Taking the P.A.$\sim 100\degr$ EofN of the semimajor axis with
redshifted velocities in NGC 5597, and assuming the optical spiral arms are trailing, the hemisphere from NW clockwise to SE is closer to the observer with the direction of the axis of rotation projected on the plane of the sky pointing SW at P.A.$\sim 190\degr$ EofN. As can be seen in Figure 4-Right, at this angular resolution and sensitivity the \hi\ 21 cm neutral cold gas is confined to the disk.

\subsection{NGC 5597 VLA B-Configuration: Neutral Atomic Gas Mass}

    The total integrated \hi\ flux is, $\int S_{\rm HI} dv \sim 2.9$ Jy
km s$^{-1}$ using the task {\it IRING} in AIPS with concentric rings, each $8\farcs0$ wide, from $R=0\farcs0 \rightarrow R=70\farcs0$ \citep{rog74}. The total estimated \hi\ 21 cm atomic hydrogen mass in NGC 5597 is M(\hi)$ \sim 1.02 \times 10^9 M_{\odot}$\footnote{The \hi\ 21 cm total mass in NGC 5597 and later in NGC 5595 was estimated from the known formula M$(\hi) = 2.356 \times 10^5 D^2 \int S_{\nu} dV$ where $D$ is in Mpc, $S_{\nu}$ in Jy and $dV$ is in km\,s$^{-1}$ \citep{wri74}}. The dynamical mass in NGC 5597 is $M_{\rm dyn} \sim 2.6 \times 10^{10}~M_{\odot}$ as measured from the observed maximum velocity; see Figure 3 Middle panel\footnote{The dynamical mass is estimated from the known formula resulting from the centrifugal force in circular orbits and the gravitational force from a central massive object, thus one has the expressions for radial forces: F$_C = mV^2/R$, F$_G=GMm/(R^2)$. The object (cloud of gas) is in equilibrium of forces, thus when we equal F$_G$ to F$_C$ one obtains the expression $M_{\rm dyn} = 233.1 V^2R$, where $V$ is in km\,s$^{-1}$, $R$ is in pc, and the mass is in solar masses.}.

\subsection{20 cm Radio Continuum and H$\alpha$ Emission from NGC 5597}

    Previous VLA 20 cm radio continuum emission observations of NGC 
5597 by \citet{con90} reported an unresolved central structure elongated at P.A.$ \sim 43\degr$ EofN, see last two values in Table 2.

    We have produced a new 20\,cm continuum emission image from our VLA
B-configuration observing sessions, shown in Figure 7-Right with a restoring beam at FWHM of $\sim 6\farcs11 \times 3\farcs7$ (P.A.$\sim -8\degr$ EofN) and contours superimposed on the optical red filter I image of NGC 5597 in grey scale \citep{gar96}. Our 20\,cm radio continuum image, consistent with previously published images at a similar wavelength \citep{con90}, shows a centrally peaked unresolved structure elongated along P.A.$\sim 15\degr$ EofN. The peak flux density of this structure is 9.02 mJy, and the rms noise in the image is 1$\sigma \sim 167.4$ $\mu$Jy\,beam$^{-1}$. Additionally, there are at least seven other weaker sources. They are all listed in Table 2 and shown in Figure 7-Right with letters from $a$ through $h$.

\begin{figure*}[thb!]
\includegraphics[width=8cm,height=8cm]{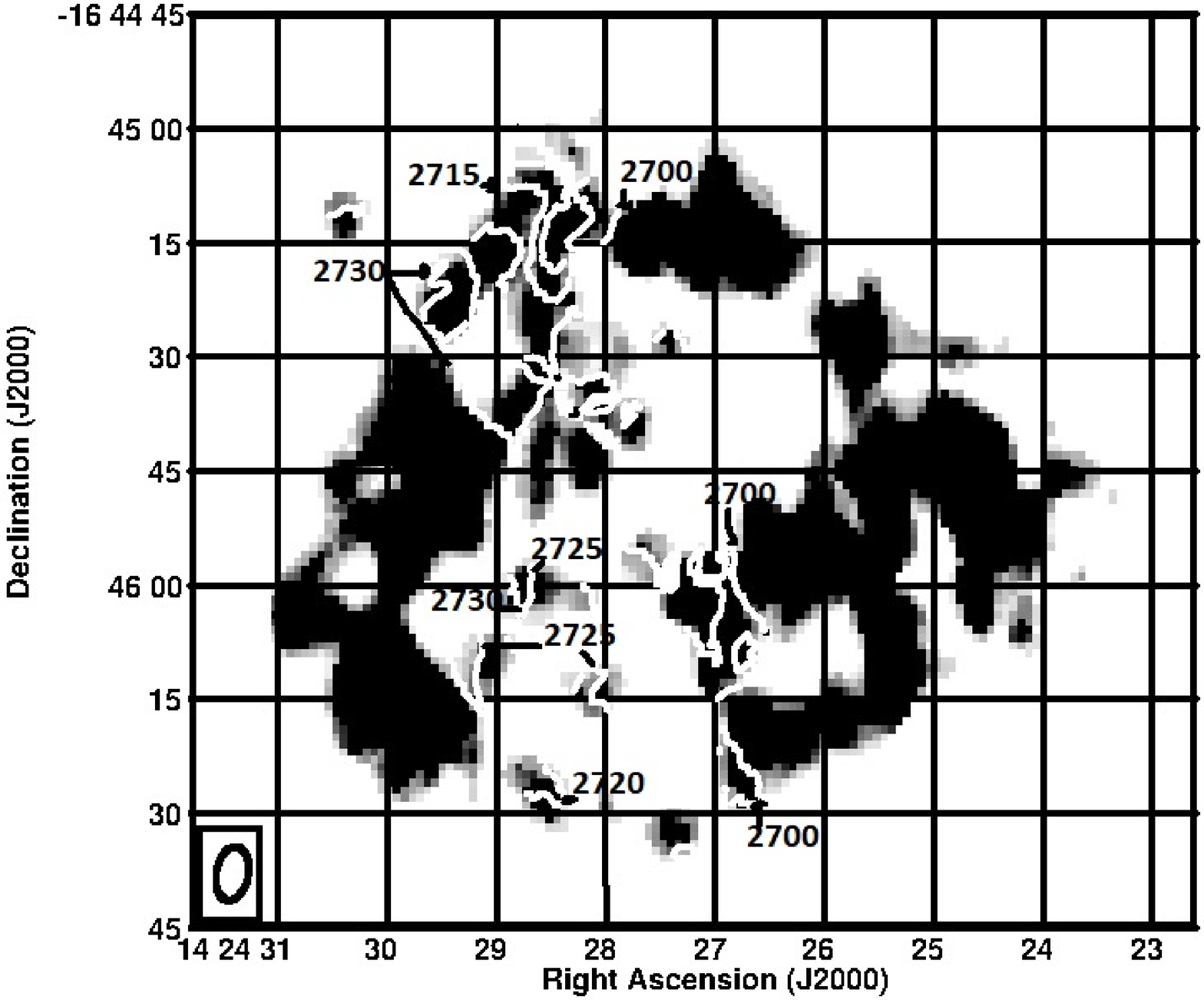}
\includegraphics[width=8cm,height=8cm]{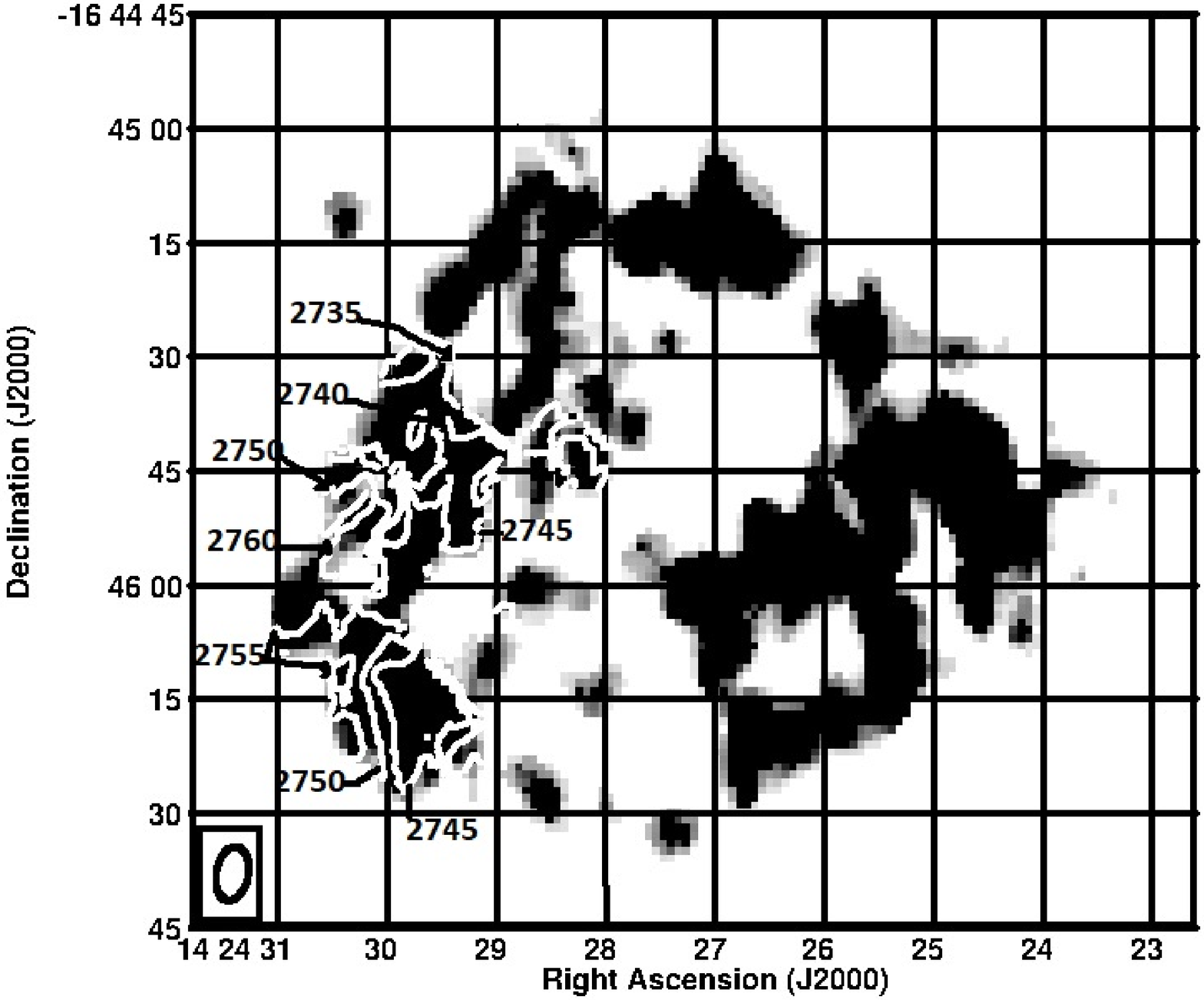}
\figcaption{Redshifted \hi\ 21 cm velocity field, Moment 1, of NGC 5597 in contours superposed on the Moment 0 grey scale image. The gray scale in both panels ranges from 1 mJy\,beam$^{-1}$\,km\,s$^{-1}$ to 15 mJy\,beam$^{-1}$\,km\,s$^{-1}$.
Left: The low redshifted velocity contours from center to left are at 2700, 2705, 2710, 2715, 2720, 2725 and 2730 km s$^{-1}$. Right: The high redshifted velocity contours from center to left are at 2735, 2740, 2745, 2750, 2755, and 2760 km s$^{-1}$.
\label{fig. 5}}
\end{figure*}

\begin{figure*}[thb!]
\includegraphics[width=8cm,height=8cm]{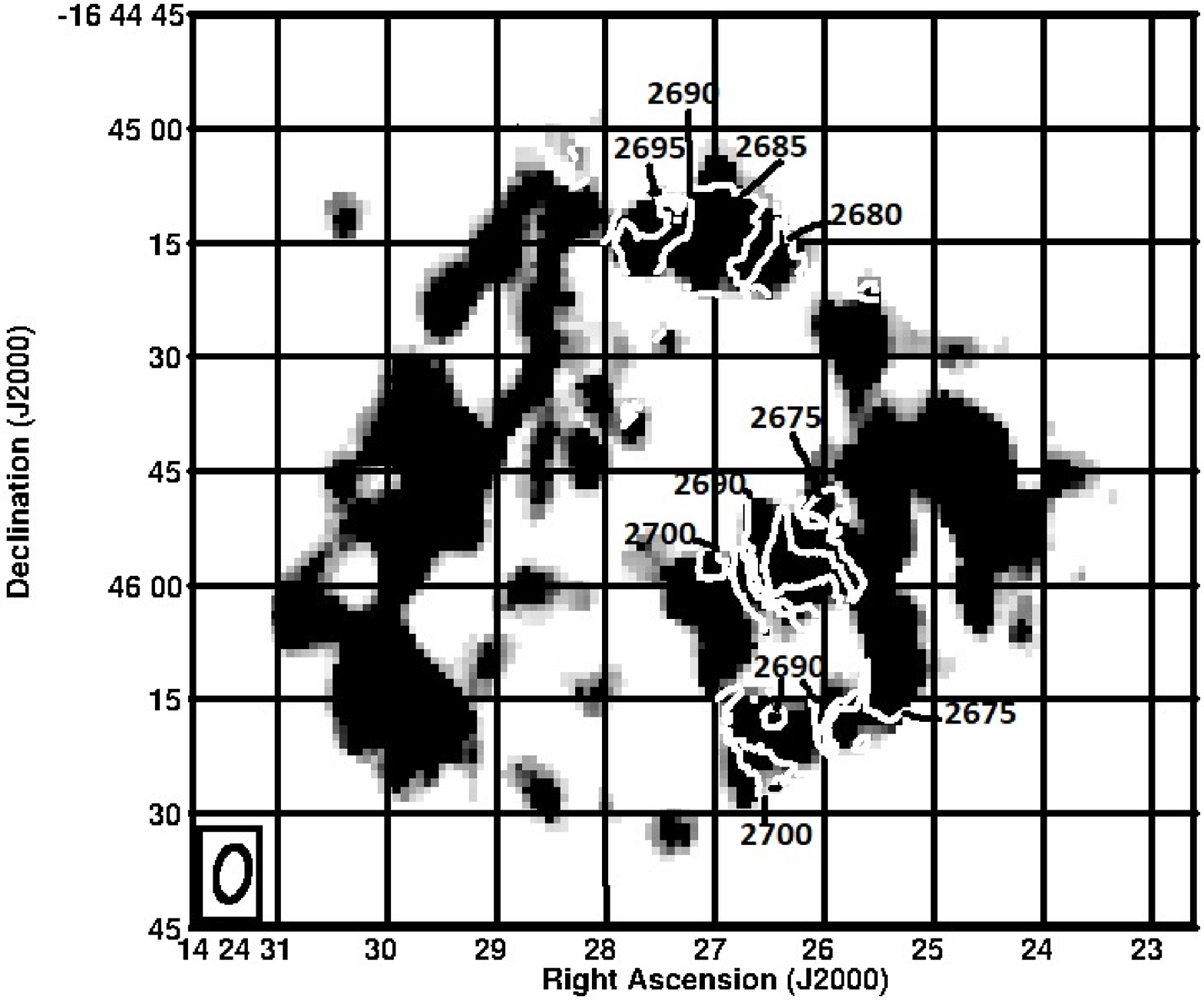}
\includegraphics[width=8cm,height=8cm]{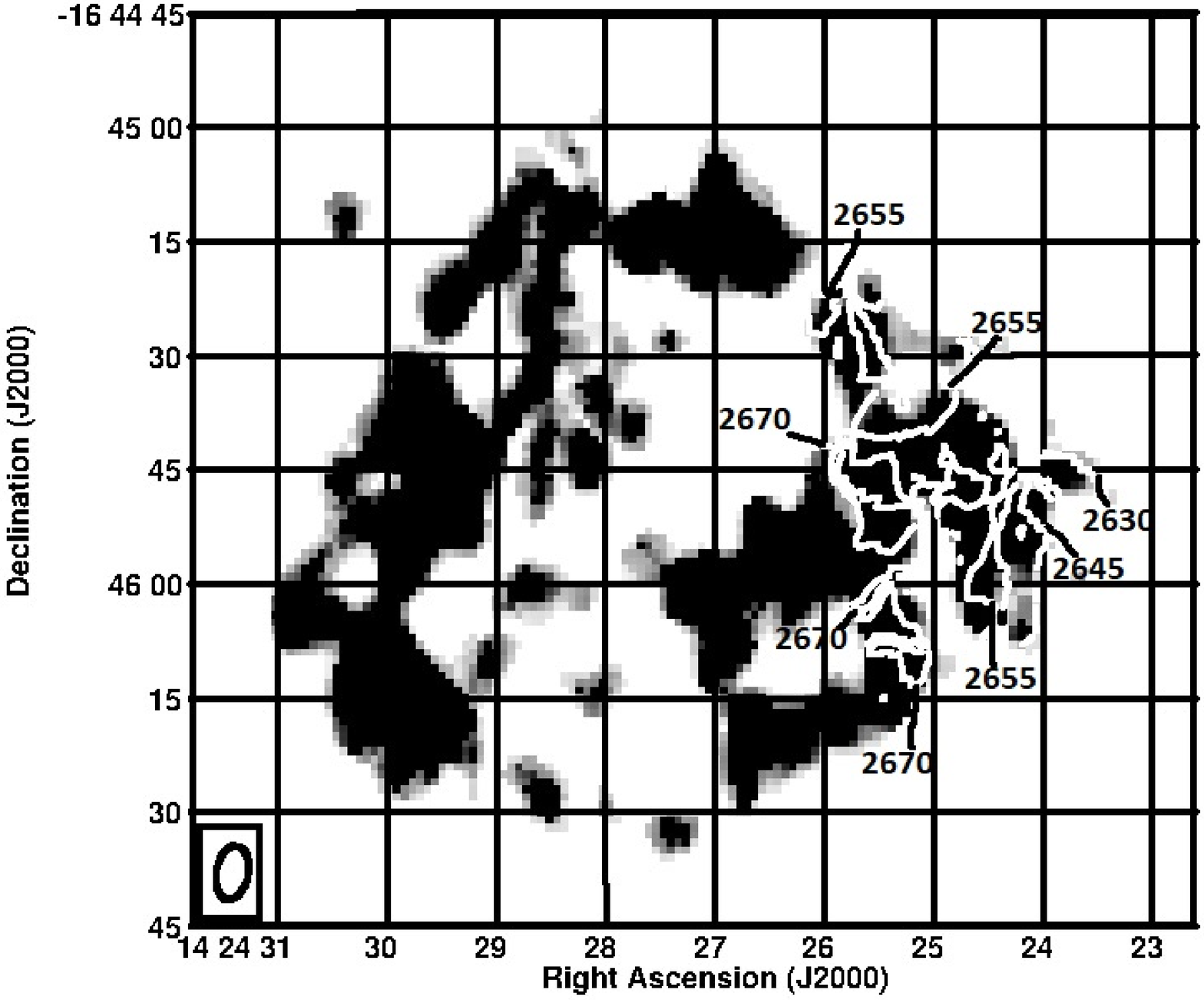}
\figcaption{Blueshifted \hi\ 21 cm velocity field, Moment 1, of NGC 5597 in contours superposed on the Moment 0 grey scale image. The gray scale ranges from 1 mJy\,beam$^{-1}$\,km\,s$^{-1}$ to 15 mJy\,beam$^{-1}$\,km\,s$^{-1}$. Left: the low blueshifted velocity contours from center to right are at 2700, 2695, 2690, 2685, 2680, and 2675 km s$^{-1}$. Right: The high blueshifted velocity contours from center to right are at 2670, 2665, 2660, 2655, 2650, 2645, 2640, 2635, 2630, 2625 and 2620 km s$^{-1}$.
\label{fig. 6}}
\end{figure*}

    The total flux density in our 20\,cm image is 37 mJy. At the
sensitivity of our observations, there is only one 20 cm radio continuum source with a peak flux density of 3$\sigma$ in the optical north-arm labeled $h$ but does not coincide with any weak H$\alpha$ source (see Figure 7-Left panel).

    Figure 7-Left shows a reproduction of our H$\alpha$+[N II]
continuum-free image of NGC 5597 \citep{gar96} in contours\footnote{H$\alpha$+[N II] image was not flux density scale calibrated, thus the contours are proportional to the rms noise in relative units.} superimposed on the \hi\ 21 cm Moment 0 image in grey scale. The most intense H$\alpha$ emission originates from an unresolved central source with its circumnuclear  region and an elongated $13''$ structure in P.A.$\sim 32\degr \pm 14\degr$ EofN. Additionally, there are at least seven unresolved H$\alpha$ sources, with low intensity emission in their surrounding associated with the inner optical eastern arm structure (labeled 1 in Figure 2), at least two in the SW optical arm (labeled 2 in Figure 2), extended low intensity emission associated with the optical arm (labeled 3 in Figure 2) and at least 9 unresolved sources associated with the north arm (labeled 4 in Figure 2). The peak of the bright H$\alpha$ emission from the north optical arm (labeled 4) is at slightly larger distance than the peak of the \hi\ 21 cm emission in that arm. 

\startlongtable
\begin{deluxetable*}{ccccc}
\tabletypesize{\scriptsize}
\tablecaption{Spatial Positions of the Nucleus in NGC 5597 measured in various Bands/Images, and the positions of other 20\,cm continuum sources in NGC 5597} \label{tab:table2}
\tablehead{
\colhead{Image} & \colhead{$\alpha$(J2000)} & \colhead{$\delta$(J2000)} & Source & Reference\\
\colhead{name}  & \colhead{$hh~mm~ss.ss$}  & \colhead{$\degr~~' ~~''$} &  & }
\colnumbers
\startdata
103aO        &  $14~24~27.49$ & $-16~45~45.9$  & nucleus & 1 \\
I            &  $14~24~27.44$ & $-16~45~45.9$  & nucleus & 2 \\
\hi\ 21cm    &  $14~24~27.16$ & $-16~45~46.64$ & nucleus & 3 \\
H$\alpha$    &  $14~24~27.34$ & $-16~45~47.39$ & peak of emission & 2 \\
20\,cm cont. &  $14~24~27.40$ & $-16~45~46.00$ & nucleus$^a$ & 3 \\
20\,cm cont  &  $14~24~28.03$ & $-16~45~48.00$ & E on disk$^b$ & 3 \\
20\,cm cont  &  $14~24~29.07$ & $-16~45~40.00$ & NE on disk$^c$ & 3 \\
20\,cm cont  &  $14~24~29.91$ & $-16~46~04.00$ & SE on disk$^d$ & 3 \\
20\,cm cont  &  $14~24~27.47$ & $-16~46~06.00$ & S on disk$^e$ & 3 \\
20\,cm cont  &  $14~24~26.36$ & $-16~45~49.00$ & W on disk$^f$ & 3 \\
20\,cm cont  &  $14~24~24.97$ & $-16~45~49.00$ & W on disk$^g$ & 3 \\
20\,cm cont  &  $14~24~28.40$ & $-16~45~07.00$ & N on disk$^h$ & 3 \\
old 20\,cm   &  $14~24~27.45$ & $-16~45~45.27$ & nucleus & 4 \\
old 20\,cm   &  $14~24~27.35$ & $-16~45~45.27$ & nucleus & 4 \\
\hline
\enddata
\tablenotetext{a}{53.9$\sigma$, $^b$ 6.7$\sigma$, $^c$ 4.2$\sigma$, $^d$ 3.8$\sigma$, $^e$ 4.2$\sigma$, $^f$ 4.3$\sigma$, $^g$ 4.1$\sigma$, $^h$ 3$\sigma$ where 1$\sigma \sim 167 \mu$Jy/beam. These peaks of the 20\,cm continuum sources are labeled
$a$ through $h$ in Figure 7-Right.}
\tablecomments{1) \citep{dia09}, 2) \citep{gar96}, 3) This work, 4) \citep{con90} }
\tablecomments{The first old 20\,cm radio continuum position by Condon et al. (1990) was with a circular beam FWHM$ \sim 21\farcs0$. The second old position was with a circular beam FWHM$ \sim 7\farcs0$. }
\end{deluxetable*}

    The lack of cold atomic \hi\ 21 cm emission in the inner $20''$
together with the presence of extended hot gas\footnote{By hot gas we mean gas from synchrotron plus thermal processes from observed H$\alpha$ and 20 cm radio continuum optically thin emissions that indicate T$_e \sim 10^4$ K.} in a P.A. similar to the P.A. of the rotation axis of the galaxy may indicate the presence of a low velocity bipolar outflow. While this extended hot gas might also be the result of central star formation, explaining its P.A. as seen in both H$\alpha$ and 20 cm radio continuum would be a challenge.

    All of H$\alpha$ sources throughout the disk indicate regions of
massive O and B star formation with ionization and recombination processes of atomic hydrogen \citep{spi78} with hot gas, $T \sim 10^4$\,K, extending out of the plane of the galaxy. The temperature of the \hi\ 21 cm gas presented in this paper, however, corresponds to an \hi\ spin temperature of $T_{\rm s} \leq 100$\,K \citep{pur56,fie58a,fie58b,wri74,spi78,com02}. Assuming the neutral gas distribution in NGC 5597 is similar to that in our galaxy, the \hi\ 21 cm cold gas and sodium, Na I, are constituents of diffuse neutral clouds in a thin layer on the plane of the galaxy's disk \citep{spi78} with thickness probably less than 200 pc \citep{bur74}.

\begin{figure*}[bht]
\includegraphics[width=8cm,height=8cm]{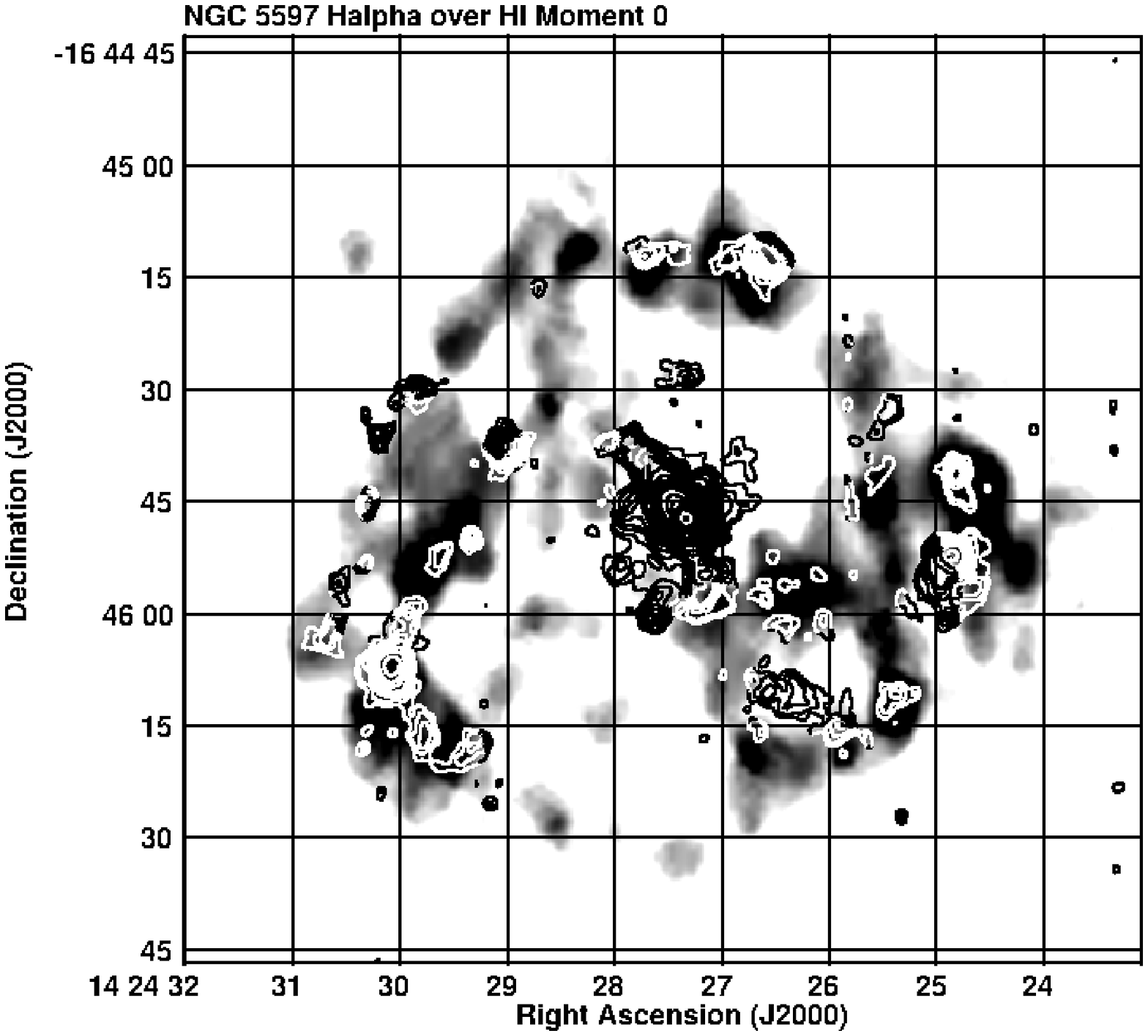}
\includegraphics[width=8cm,height=8cm]{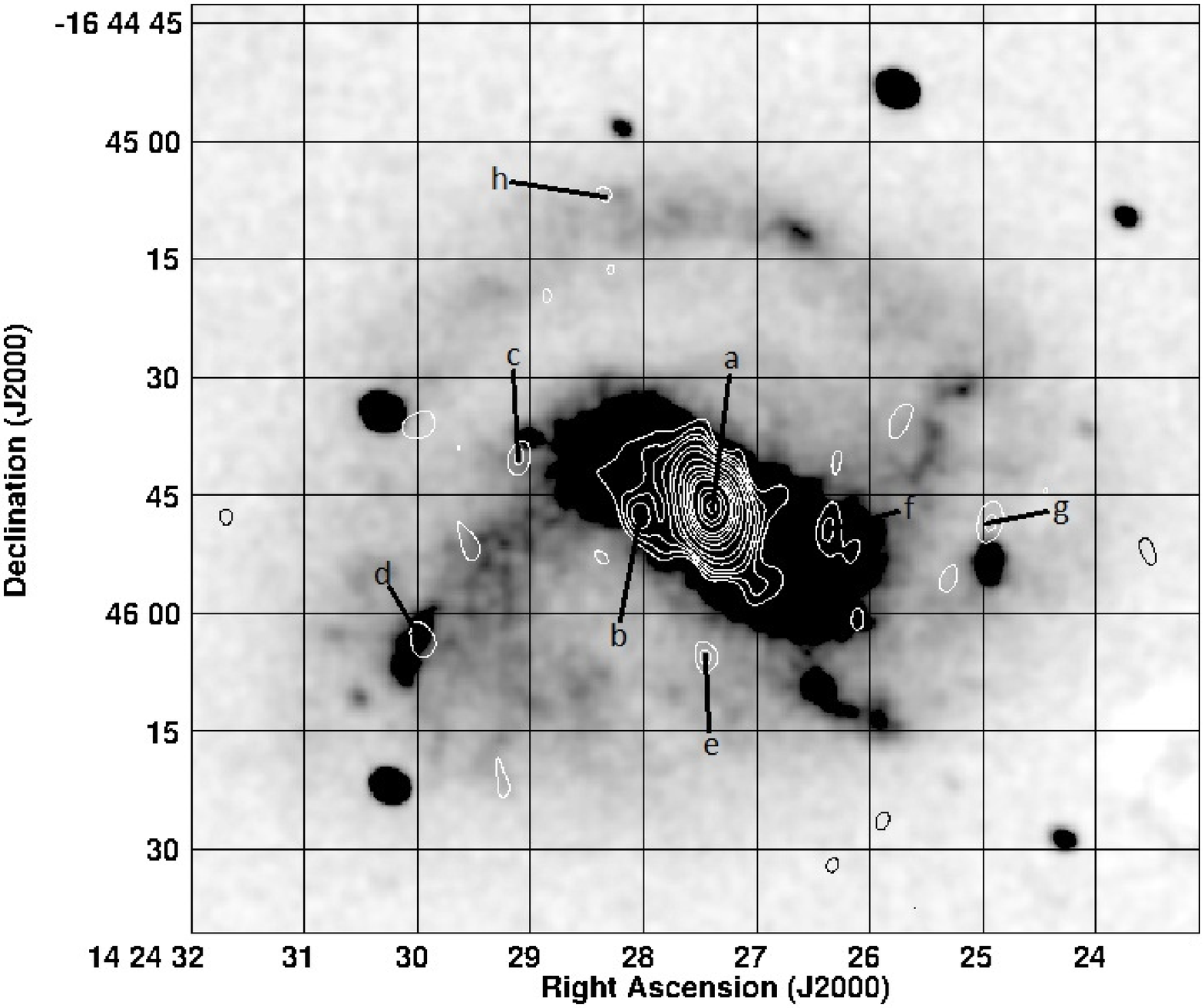}
\figcaption{Left: reproduction of the H$\alpha$ line emission image of NGC 5597 \citep{gar96} in contours superimposed on the \hi\ 21 cm Moment 0 in grey scale. The H$\alpha$ image is not calibrated in flux density scale, therefore the contours are proportional to noise in arbitrary units from 3.5$\sigma$ to 17$\sigma$. The most intense central emission of the hot gas, seen in H$\alpha$, is devoid of cold gas \hi\ 21 cm emission. At least two H$\alpha$ sources are offset by a few arcsec to the north of the \hi\ 21 cm sources. Right: Our new 20 cm radio continuum emission image in contours superimposed on the optical red (I filter 8040 \AA) image \citep{gar96}. The restoring beam size of the 20 cm continuum image at FWHM is $\sim 6\farcs11 \times 3\farcs7$ (P.A.$\sim -8\degr$). The contours are from 3, 4,, 5, 6, 8, 10, 14, 18, 22, 26, 30, 40, 45 and 53$\sigma$ where 1$\sigma \sim 167.4$ $\mu$Jy\,beam$^{-1}$. The strongest 20\,cm radio continuum emission arises from an unresolved source coincident with the position of the optical nucleus with an inner elongated structure at a P.A.$\sim 25\degr$ EofN. The peaks of 20\,cm continuum sources are labeled $a$ through $h$ as listed in Table 2.
\label{fig. 7}}
\end{figure*}

\begin{figure*}[bht]
\includegraphics[width=8cm,height=8cm]{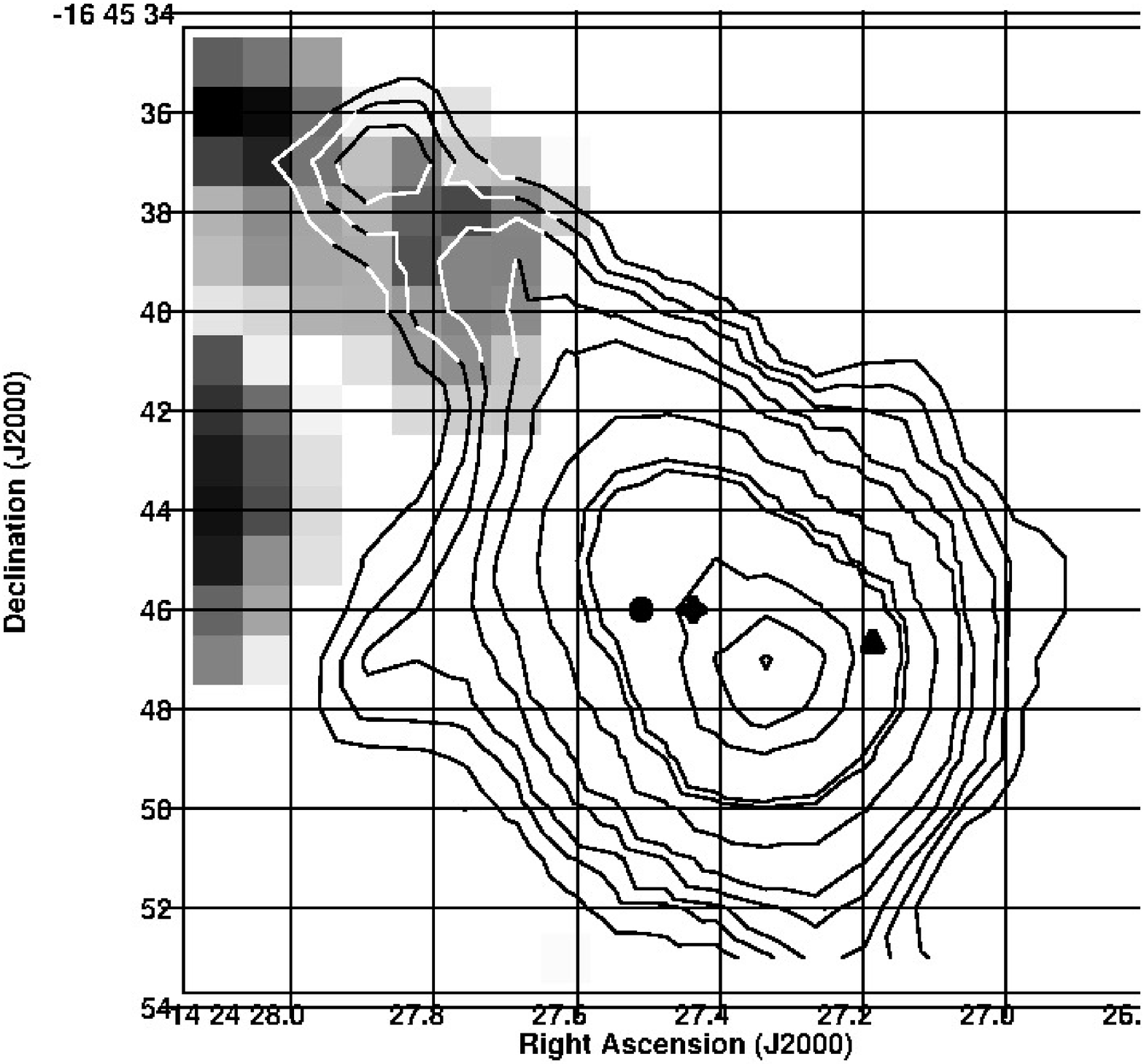}
\includegraphics[width=8cm,height=8cm]{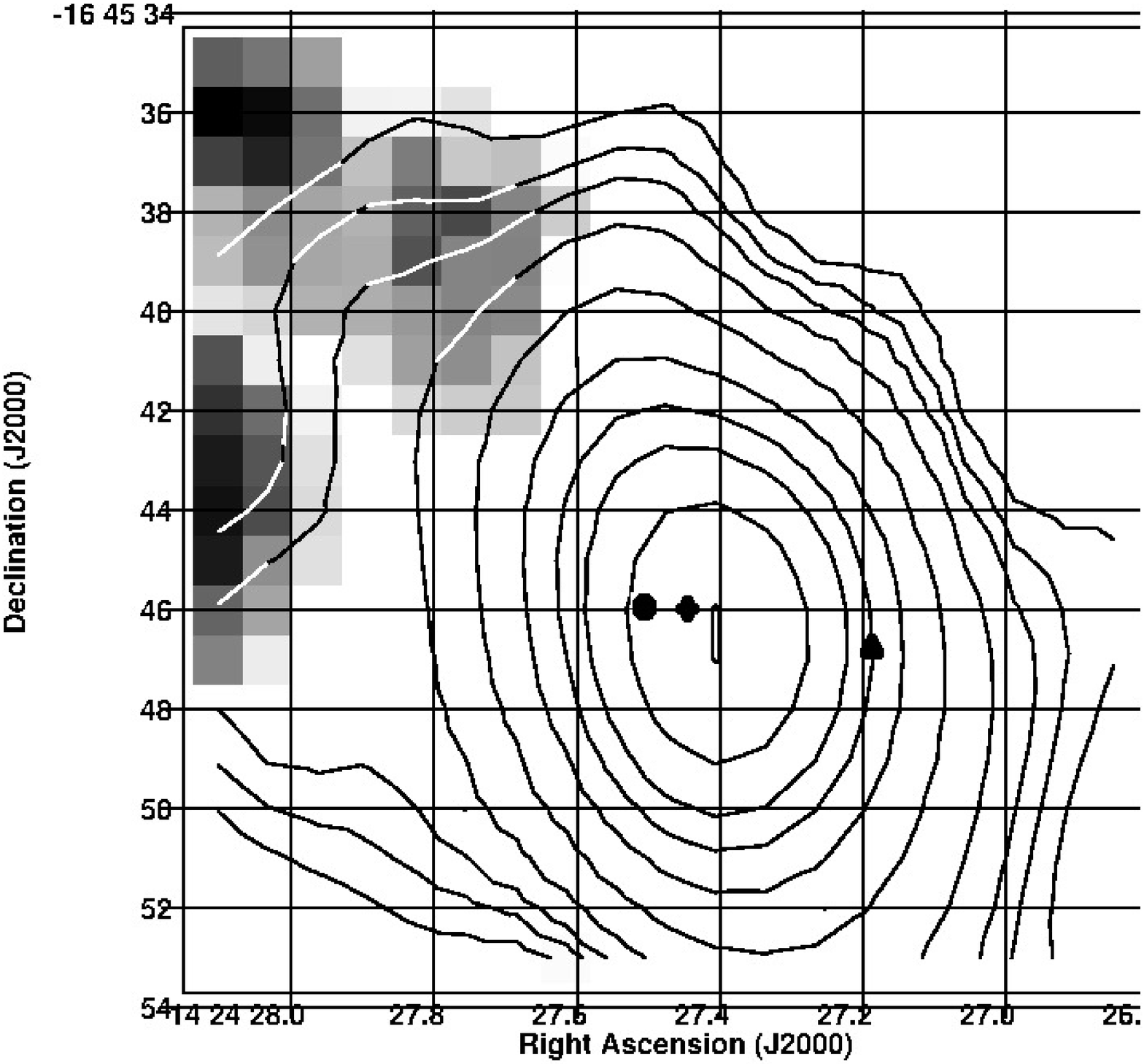}
\figcaption{Innermost $20''$ of NGC 5597, the filled triangle shows the kinematical center of \hi\ 21 cm emission, the four point star shows the position of the photometric optical red I filter image, and the filled circle shows the position of the photometric blue 103aO image. Spatial coordinates are listed in Table 2. Left: H$\alpha$ continuum-free emission in contours superposed on VLA B-configuration \hi\ 21 cm Moment 0 image in grey scale. The optical image is not calibrated in flux density scale, and the contour levels are at 7, 9, 11, 15, 20, 40, 80, 100, 300, 500, and 720 times 1$\sigma$ where 1$\sigma = 10$ in  relative units. Grey scale ranges from 5.76 to 30 mJy\,beam$^{-1}$\,km\,s$^{-1}$. The extended emission lies at P.A.$\sim 32\degr \pm 14\degr$. Right: our new VLA B-configuration 20 cm continuum emission in contours superposed on the \hi\ 21 cm MOM 0 image in grey scale. The contour levels are the same as in Figure 8b. The NE extended 20 cm radio continuum is in P.A.$\sim 17\degr \pm 7\degr$. Grey scale ranges from 5 to 25 mJy\,beam$^{-1}$\,km\,s$^{-1}$. Notice that the peak of the H$\alpha$ emission (left panel) is slightly south-west from the peak of the 20 cm radio continuum (right panel) that is associated with the optical nucleus. Qualitatively both the 20 cm continuum and the H$\alpha$ spatial emissions are similar. There is no cold gas atomic \hi\ 21 cm emission (in grey scale) from the nucleus and the innermost circumnuclear region. 
\label{fig. 8}}
\end{figure*}

\begin{figure*}[bht]
\includegraphics[width=6cm,height=6cm]{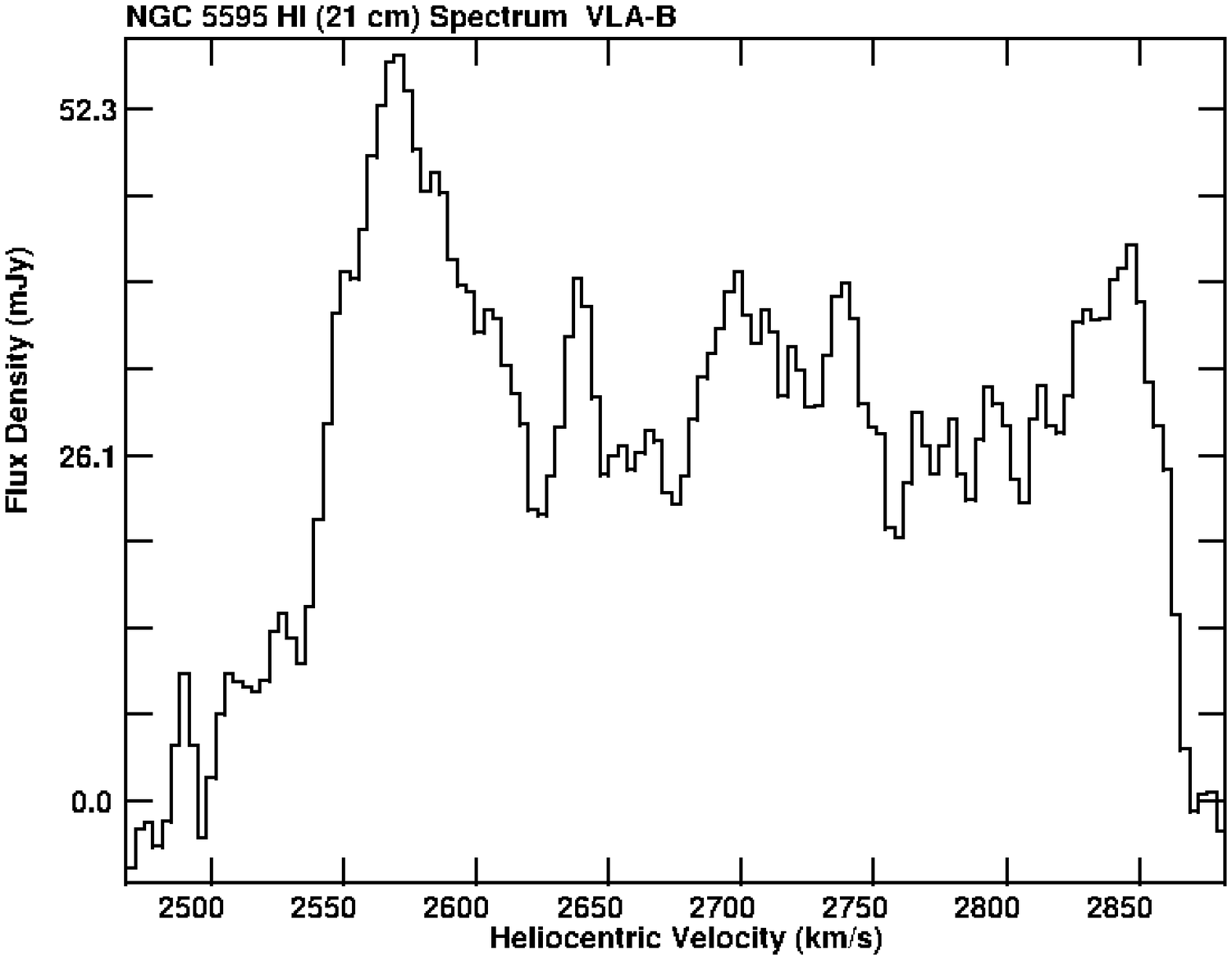}
\includegraphics[width=6cm,height=6cm]{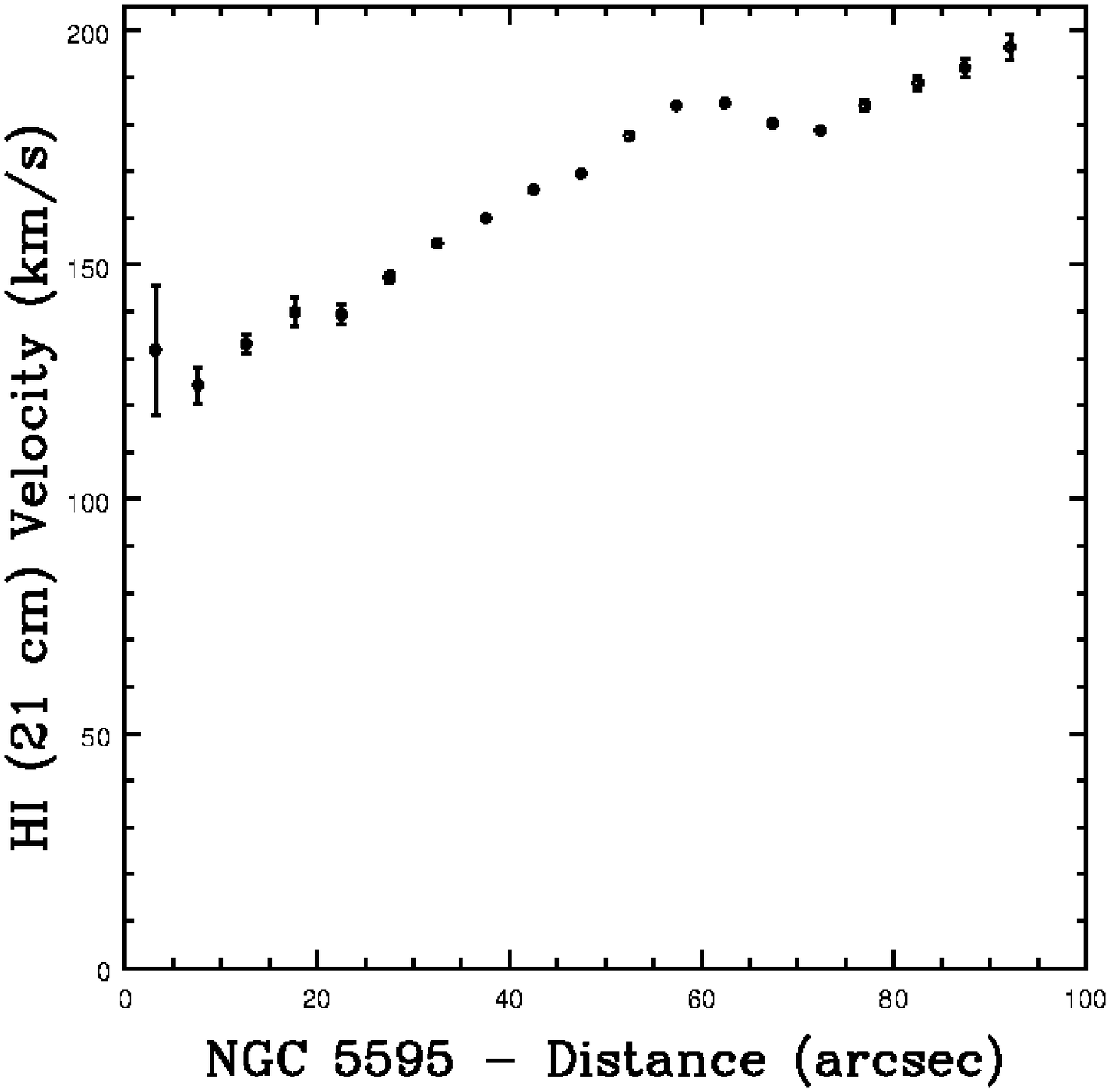}
\includegraphics[width=6cm,height=6cm]{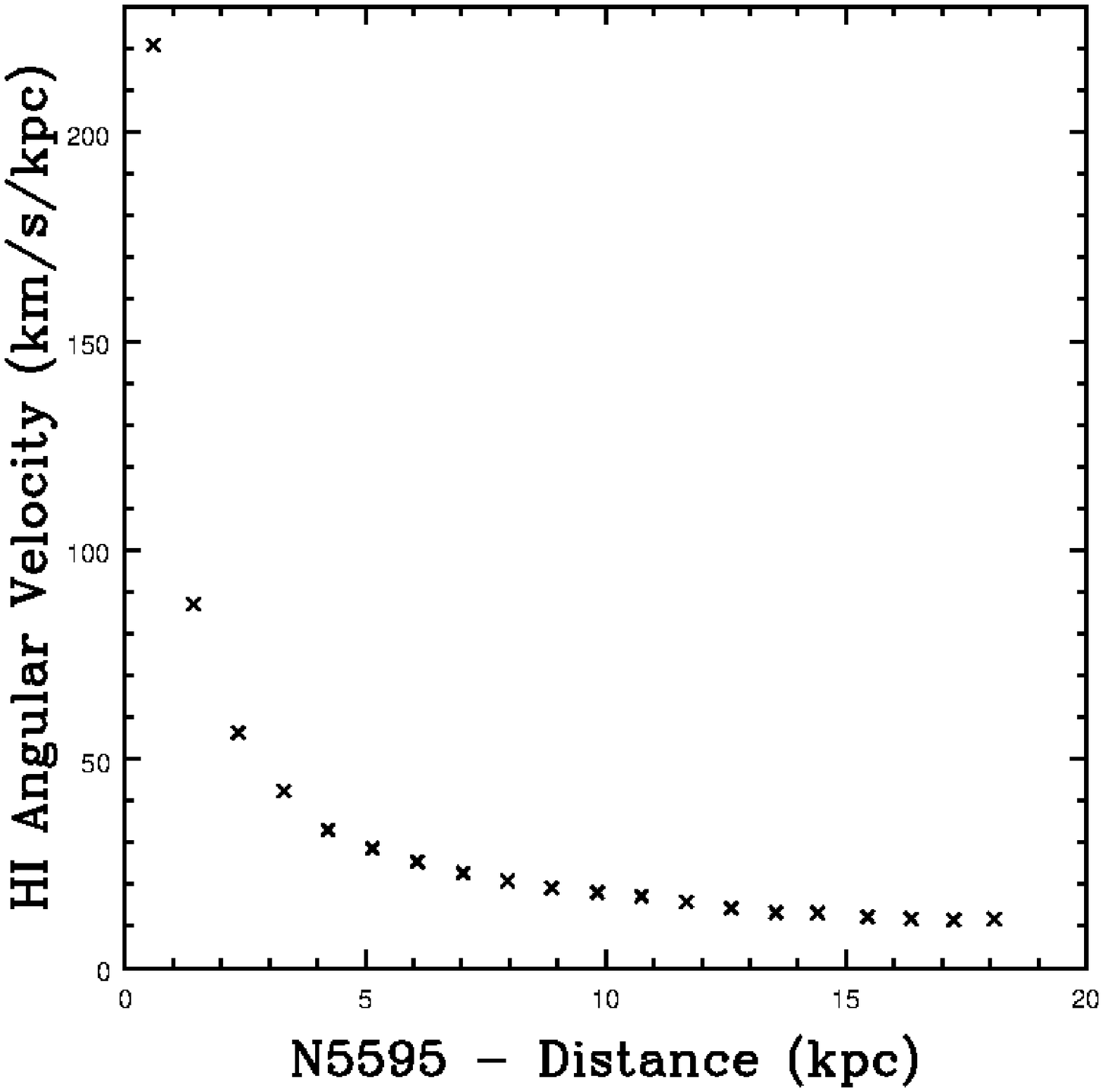}
\figcaption{Left: \hi\ 21 cm spectrum obtained from our VLA B-configuration observations of the disk galaxy NGC 5595. The heliocentric systemic velocity, fitted by the task {\it GAL} in AIPS, is $V(\rm sys)_{\rm helio} = 2702$ km s$^{-1}$, with $\Delta V_{50\%} \sim 317.5$ km s$^{-1}$, and $\Delta V_{20\%} \sim 345$ km s$^{-1}$. The flux density scale of the spectrum is very difficult to compare to previous ones obtained with Parkes (64 m) and Green Bank (91m) radio telescopes because their beam included both disk galaxies NGC 5595 and NGC 5597 \citep{mat92,spr05}. The shape of the spectrum looks very similar to that obtained with Nan{\c{c}}ay, however its beam at FWHM was $\sim 3\rlap{.}{'}6$ east-west $\times~22'$ north-south \citep{pat03}. Middle: NGC 5595 \hi\ 21 cm rotation curve. Right: NGC 5595 \hi\ 21 cm angular velocity $\Omega_{\rm gas}$. \label{fig.9}}
\end{figure*}

\begin{figure*}[bht]
\begin{center}
\includegraphics[width=12cm,height=12cm]{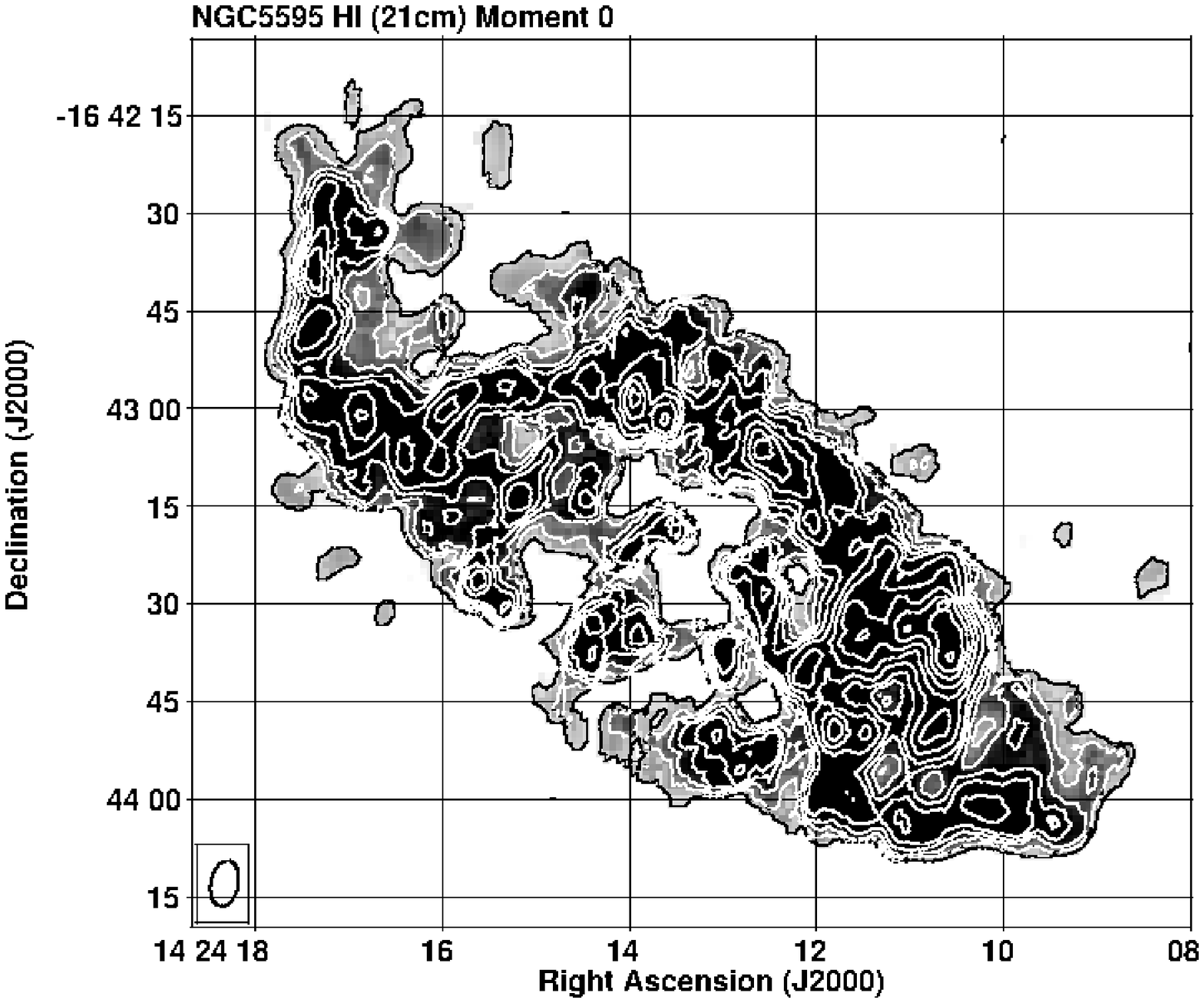}
\figcaption{NGC 5595 VLA B-configuration \hi\ 21 cm integrated intensity over velocity, Moment 0, image in grey scale and contours. The contour levels are at 3, 6, 9, 12, 15, 20, 25, 30, 35, 40 and 47 $\times~1\sigma = 3.3$ Jy\,beam$^{-1}$\,km\,s$^{-1}$. Notice that there is no atomic hydrogen gas emission from the position of the kinematical center $\alpha$(J2000.0) = $14^{\rm h}24^{\rm m}13\rlap{.}^{\rm s}3$, $\delta$(J2000.0) = $-16\degr43'21\farcs6$. This position coincides with the optical continuum photometric center. The VLA B-configuration beam size, which is $\sim 7\farcs14 \times 4\farcs21$ at FWHM, is shown in the lower left corner.  \label{fig. 10}}
\end{center}
\end{figure*}

\subsection{Central $20''$ Gas Emission from NGC 5597}

    Figure 8-Right shows the central innermost $\sim 20''$ image of
our new 20 cm continuum emission in contours obtained with the VLA in B-configuration. The image is superposed on the grey scale image of the \hi\ 21 cm Moment 0 of NGC 5597. The peak of the 20 cm continuum emission at this angular resolution (FWHM of $\sim 6\farcs11 \times 3\farcs7$, P.A. $\sim -8\degr$) coincides fairly well with the position of the nucleus seen in the blue optical 103aO and the red optical (I filter) images. Table 2 list the positions of the center or nucleus\footnote{\hi\ 21 cm kinematic, photometric 103aO and I optical filters.} in NGC 5597 and they are also shown with different symbols in Figure 8.

    The new 20 cm radio continuum map shows that the emission arises
from an unresolved source with an elongated structure in the NE-SW direction at P.A.$\sim 17 \pm 7\degr$ EofN, This elongation is very comparable to the galaxy's rotation axis projected on the plane of the sky (direction P.A.$\sim 190\degr$ EofN or orientation P.A.$\sim 10\degr$ EofN). This NE-SW 20 cm radio continuum elongation is a true physical orientation and not due to a beam elongation artifact (since the beam P.A.$\sim -8\degr$ EofN). Also, it is in a similar orientation as the elongated structure seen in H$\alpha$, which we discuss further below. 
 
    The observed 20 cm continuum emission is a mixture of synchrotron
process with high velocity electrons interacting with an external magnetic field, $\vec{B}$\footnote{For optically thin synchrotron emission, $S_{\rm \nu} \propto B_{\perp}^{1.75} \nu^{-0.75} \theta^3$, while the free-free emission for an optically thin case, $S_{\nu} \propto T_e^{-0.35} \nu^{-0.1}EM$.}, and thermal process (free free optically thin) with a typical electron temperature of $T_e \sim 10^4$\,K from \hii\ regions and/or supernova processes\footnote{Future VLA radio continuum observations of NGC 5597 at a higher angular resolution and polarization analysis would be critical to better understand the physical mechanisms and the $\vec{E}$ (and $\vec{B}$) orientation.}.

   Figure 8-Left shows in contours the H$\alpha + $[N II] continuum-
free line emission, superposed on the grey scale VLA B-configuration \hi\ 21 cm Moment 0 map. Similar to the 20\,cm radio continuum emission, the H$\alpha$ emission in the center of the galaxy originates from an unresolved source with an elongated structure in the NE-SW direction at P.A.$ \sim 32 \pm 14\degr$ or P.A.$ \sim 212 \pm 14\degr$. This H$\alpha$ emission comes from a recombination line process in the innermost \hii\ circumnuclear region.

    Both the central H$\alpha$ and 20 cm radio continuum emissions are
elongated in the NE - SW direction and qualitatively their spatial extension is similar.  The stellar bar (in the optical red, I, filter) lies at P.A.$\sim 52\degr$ and rotates at $\Omega_{\rm bar} \sim 15.3$ km s$^{-1}$ kpc$^{-1}$ \citep{gar96,gar22}. In that central region there is no \hi\ 21 cm emission. Furthermore, the spatial distribution of both the hot gas, seen in H$\alpha$, and the thermal/synchrotron 20 cm radio continuum suggests a bipolar geyser\footnote{The term `geyser' is utilized as a synonym to low velocity (hundreds of km\,s$^{-1}$) extragalactic nuclear outflow \citep{moo95}. In contrast, extragalactic plasma outflows, like in M87, have much higher velocities (many thousands of km\,s$^{-1}$)\citep{rei89}.} into the innermost NE -- SW direction with additional emission from an innermost circumnuclear region. 

    In both images in Figure 8 the \hi\ 21 cm emission, shown in grey
scale, is only seen from the extreme NE of the central $20''$ region of NGC 5597. The peak of the H$\alpha$ emission lies slightly south-west of the optical photometric position of the nucleus (see Table 2). Although the small difference in these two positions might be due to astrometry, it might well be the correct spatial position of the SW H$\alpha$ emission (brightest) from a bipolar nuclear geyser as observed clearly in the barred disk galaxy NGC 1415 where there is weak or no H$\alpha$ emission from the optical nucleus but two bright unresolved H$\alpha$ sources from the lobes of a geyser \citep{gar96,gar00,gar19}. In this scenario, the NE lobe of such an H$\alpha$ bipolar geyser or outflow in NGC 5597 might also exist, however, our image does not have the proper angular resolution to reveal it, especially that the emission is mixed with the elongated structure in the NE-SW direction (see Figure 7-left).

    If the H$\alpha$ and the 20 cm radio continuum emission from NGC
5597 do indeed indicate a bipolar geyser or nuclear outflow, the hot ionized gas and the relativistic electrons would be out of the plane of rotation of this galaxy, similar to e.g., the large lobes seen in radio continuum in the barred galaxy NGC 3367 \citep{gar98,gar02}.

    Although we are not yet aware of any published work on molecular gas
emission in NGC 5597, one may expect a total M$(H_2) \sim 6.4 \times 10^9 M_{\odot}$ from the CO - FIR correlation \citep{sco88a}. For example, molecular gas emission has been observed from the innermost central regions of two nearby bright barred galaxies like NGC 1068 \citep{pla91} where there is a nuclear bipolar radio continuum emission \citep{van82,wil87} and NGC 3367 \citep{gar98,gar02,gar05}.

    From the literature, there are many normal disk or barred galaxies
with weak active galactic nuclei, such as the following four examples: a) M51, SA(s)bc pec, is a normal disk galaxy with a central mild activity that indicates the presence of a bidirectional jet, shown by H$\alpha$ + N[II] monochromatic image, optical red spectra, and radio continuum emission associated with the central region emanating from the nucleus as seen in high angular resolution VLA observations \citep{for85,cec88,cra92}. Furthermore, there is dense molecular gas in the inner $\leq 50$ pc of M51 located in two blobs at a P.A.$ \sim 75\degr$, perpendicular to the elongation of the optical H$\alpha$, and radio continuum images \citep{sco98}; b) NGC 1068, Sb(rs) II \citep{san81}, with a stellar bar \citep{sco88a}, central molecular gas \citep{pla91}, and a bipolar radio continuum nuclear source \citep{wil87}; NGC 1415, SBa \citep{san81} with two bright H$\alpha$ unresolved sources straddling the nucleus labeled A to the south-east and B to the north-west \citep{gar19} which may the result of a bipolar geyser with $v_{rm gayser} \sim 140$ km s$^{-1}$ and d) NGC 3367, SBc(s) II \citep{san81}, with central molecular gas \citep{gar05}, central H$\alpha$ emission \citep{gar96}, and a bipolar radio continuum source \citep{gar98,gar02}. 

    If one takes the velocity dispersion of stars in the innermost
central region of each galaxy from the published literature \citep{ho09}, and use the $M_{\rm BH} - \sigma_*$ correlation relation \citep{geb00,mer01a,mer01b}, then the mass of the super massive black hole (SMBH) in each of these three barred galaxies are: $M_{\rm NGC1068-BH} \sim 9.7 \times 10^7 M_{\rm \odot}$, $M_{\rm M51-BH} \sim 7.6 \times 10^6 M_{\rm \odot}$, and $M_{\rm NGC 3367-BH} \sim 1.4 \times 10^6 M_{\rm \odot}$. Thus, the observed central H$\alpha$ emission, the 20 cm radio continuum emission, and the central \hi\ 21 cm hole, in all of these barred galaxies may indeed be the result of a central SMBH. 

    As a comparison, the masses of the SMBH in two normal disk galaxies
observed with nuclear geysers or bipolar outflows, namely M81 (NGC 3031; Sb(r)I-II) and M101 (NGC 5457, Sc(s) I), using the same $M_{\rm BH} - \sigma_*$ correlation, are: $M_{\rm M81-BH} \sim 5.4 \times 10^7 M_{\rm \odot}$ and $M_{\rm M101-BH} \sim 3.97 \times 10^4 M_{\rm \odot}$, respectively \citep{gar19}.

    In the case of the barred galaxy NGC 5597, one of the two galaxies
in this study, the central observed distribution of the H$\alpha$ emission, the 20 cm radio continuum emission, and the \hi\ 21 cm hole, all seem to suggest the existence of a SMBH with $M_{\rm NGC 5597-BH} \sim 10^6 M_{\rm \odot}$.

\section{NGC 5595: a late type Disk Galaxy}
\subsection{General Characteristics}

    NGC 5595 is classified as Sc(s)II \citep{san81},  SAB(rs)c
\citep{dev93}. It is a member of a close pair with NGC 5597 to the SE at a projected angular distance on the plane of the sky of $\sim 3\rlap{.}{'}97$ \citep{gar03}. As mentioned earlier in the case of NGC 5597, they both have very similar systemic velocities and are close on the plane of the sky, therefore we also adopt its distance as $D_{{\rm pair}} = 38.6$ Mpc \citep{tul88} ($1'' \sim 187.14$ pc). Figure 1 shows its blue continuum optical emission from our 103aO observation \citep{dia09}. As in the case of NGC 5597, astrometry was done to independently estimate its photometric position \citep{dia09}. See Table 4 for its basic properties.

\subsection{VLA B-Configuration \hi\ 21 cm Observations}

    As previously discussed in the section on NGC 5597, the Parkes and
Green Bank single dish radio telescopes did not have enough angular resolution to separate the two galaxies, resulting in spectra that showed the combined \hi\ 21 cm emission from both NGC 5595 and NGC 5597 \citep{mat92,spr05}. Only the Nan\c{c}ay radio telescope, with a FWHM beam size of $\sim 3\rlap{.}{'}6$ east-west $\times 22'$ north-south, was able to isolate the \hi\ 21 cm emission spectrum of NGC 5595 with a peak flux density of $\sim 100$ mJy and a velocity range $\sim 2525 \rightarrow 2875$ km s$^{-1}$ \citep{pat03}.

    Figure 9-Left panel shows our VLA B-configuration \hi\ 21 cm
spectrum from NGC 5595 using task {\it ISPEC} in AIPS. The shape is very similar to the Nan\c{c}ay spectrum \citep{pat03}. Using the AIPS task {\it GAL}, the fitted kinematic \hi\ 21 cm systemic velocity is $V_{\rm sys} \sim 2702$ km s$^{-1}$ using both redshifted and blueshifted velocities. The measured velocity widths in the spectrum obtained from our VLA B-configuration observations are $\Delta V_{\rm 50\%} \sim 317.5$ km s$^{-1}$ and $\Delta V_{\rm 20\%} \sim 345$ km s$^{-1}$ at 50\% and 20\%, respectively. Figure 9-Middle panel shows the rotation curve from NGC 5595 from our fit assuming circular orbits \citep{rog74}. It has a high value from a small radius then decreases a little bit then continues slowly rising all the way until 90$''$. Finally, Figure 9-Right panel shows the NGC 5595 angular velocity ($\Omega_{\rm gas} \equiv V(R)/R$); notice its very smooth decreasing values.

\subsection{\hi\ 21 cm Spatial Distribution and Kinematics}

    Figure 10 shows the VLA B-configuration \hi\ 21 cm velocity
integrated Moment 0 map from NGC 5595 in contours superimposed on itself on grey scale. Most of the emission originates from giant \hi\ 21 cm clouds in normal differential rotation around the nucleus of the disk galaxy. The spatial extent of the \hi\ 21 cm emission spans $\Delta {\rm R.A.} \sim 18^s \rightarrow 09^s$ and $\Delta {\rm Dec} \sim -16\degr42'15'' \rightarrow -16\degr44'10''$. Notice that the optical north spiral arm in NGC 5595 is assumed to be trailing (e.g., Figure 1). However, the NE \hi\ 21 cm extended morphology, on the plane of the sky, appears as if it were a leading structure\footnote{This could be a plausible result of a flat retrograde parabolic passage of a companion of equal mass, see Fig. 1 t = 4 in \citet{too72}.}.

    We also note the lack of \hi\ 21 cm emission from the central
region, where there seems to be a circumnuclear radio continuum structure (see subsection 5.5).

    Figure 11 shows our VLA B-configuration \hi\ 21 cm velocity field.
The kinematical parameters are listed in Table 3. The Left panel shows the contours of the blueshifted velocities (compared to systemic velocity), while the Right panel shows the contours of the redshifted velocities from center to SW. The NE $\rightarrow$ N $\rightarrow$ SW hemisphere is closer to the observer, and the projection on the plane of the sky for the orientation of the rotation axis of the disk galaxy is at P.A.$ \sim 327$ EofN, with a direction pointing to the NW. 

    Figure 10 reveals (as well as Figures 11 and 12) for the first time
the existence of extended \hi\ 21 cm cold gas structures to the NE and SW of the disk galaxy NGC 5595 with no blue optical continuum nor 20 cm radio continuum counterparts.The approximate length of the \hi\ 21 cm tails at the NE and the SW of NGC 5595 on the plane of the sky are $\sim 7$ kpc and $\sim 4.2$ kpc respectively. The NE contours show blueshifted velocities as if they were a smooth continuation of the blueshifted velocities from the inner disk. Similarly, the lower-most SW contours show redshifted velocities as if they were a smooth continuation of the redshifted velocities from the inner disk. 
Both (NE \& SW) extended structures seem to be warps of the NGC 5595 disk\footnote{The large scale warps of disk galaxies remain a challenge for theorists \citep{too83}.}.

    Figure 12 shows the reproduction of the blue optical continuum 103aO
emission from NGC 5595 in contours \citep{dia09} superposed on the \hi\ 21 cm Moment 0 map in grey scale (see Figure 10). The blue optical image is not flux calibrated, so the contours are proportional to 1$\sigma$ noise level in arbitrary units.

    As mentioned earlier, the NE \hi\ 21 cm extension morphology is
peculiar since it shows blueshifted velocities and it appears, on the plane of the sky, as if it were a leading structure; the north optical (blue) spiral arm is assumed to be trailing and shows also blueshifted velocities. The NE and SW \hi\ 21 cm elongated emissions may be filamentary gas structures that are most likely a result of a recent gravitational tidal interaction with its neighbor NGC 5597. Similar extended structures have been observed in the M82 galaxy \citep{yun93}, and have been shown to exist from computer simulations \citep{too72,mih96}. 

\begin{figure*}[bht]
\includegraphics[width=8cm,height=8cm]{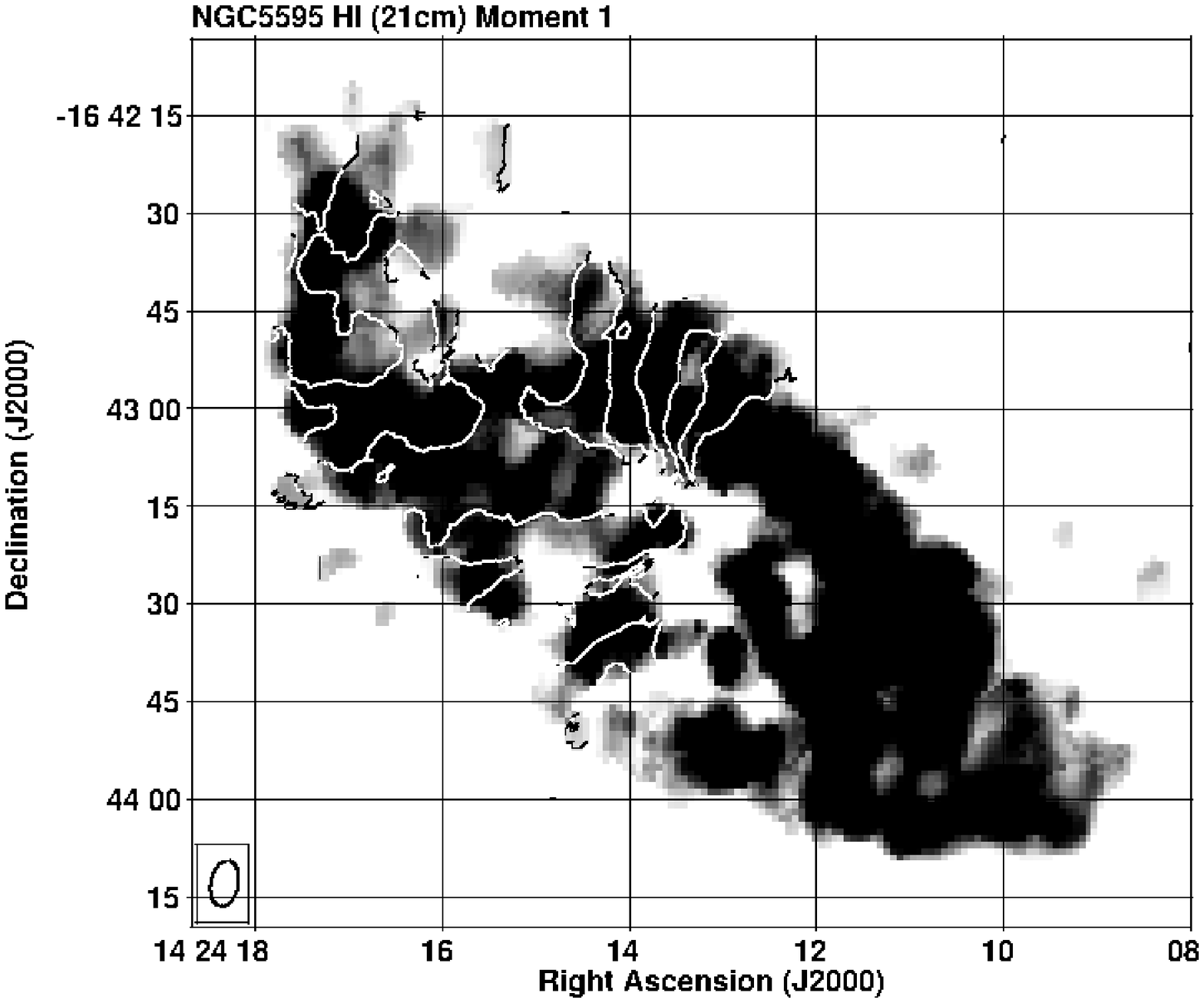}
\includegraphics[width=8cm,height=8cm]{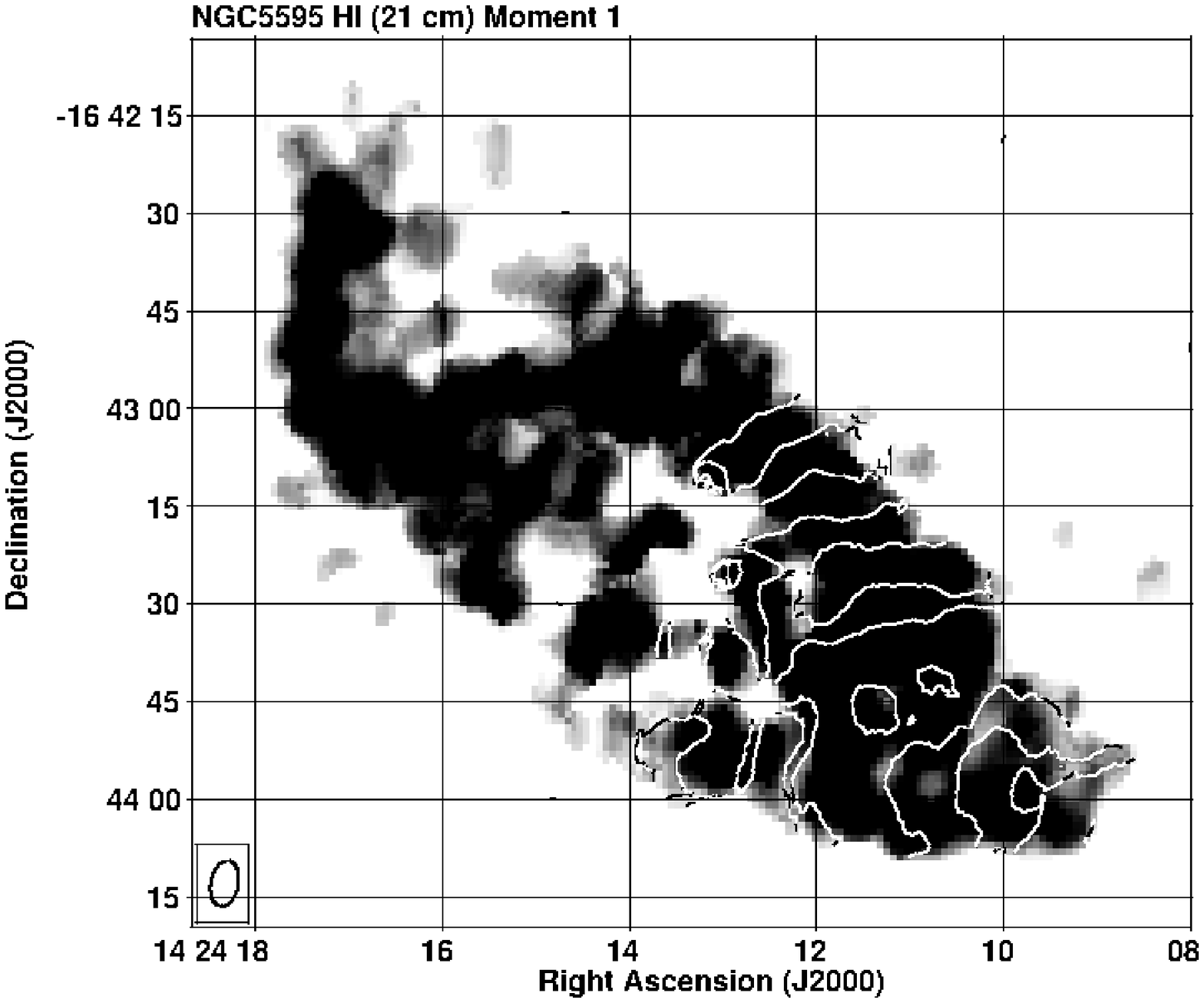}
\figcaption{Left panel shows the \hi\ 21 cm velocity field of NGC 5595, MOM 1, with blueshifted velocities in contours superposed on Moment 0 image in grey scale (scale stretch is from 7.7 mJy beam$^{-1}$ km s$^{-1}$ to 37 mJy beam$^{-1}$ km s$^{-1}$). Velocity contours from center to NE are at 2690, 2670, 2650, 2630, 2610, 2590, 2570, 2560 and 2550 km\,s$^{-1}$. Right panel shows the redshifted velocity contours from center to SW are at 2700, 2720, 2740, 2760, 2780, 2800, 2820, 2840, 2850 and 2855 km\,s$^{-1}$. The velocity field indicates differential rotation as expected from a disk galaxy. Notice that the upper-most NE contours show blueshifted velocities as if they were a smooth continuation from the blueshifted velocities from the inner disk, although the cold atomic \hi\ 21 cm gas there has no optical counterpart. Similarly the SW contours show redshifted velocities as if they were a smooth continuation from the redshifted velocities from the inner disk, although the cold atomic \hi\ 21 cm gas there has no optical counterpart.
\label{fig. 11}}
\end{figure*}

\subsection{NGC 5595: Cold Hydrogen Atomic Gas Mass}

    The total \hi\ 21 cm gas mass in NGC 5595 is $M({\rm HI})~\sim 2.9
\times 10^9 M_{\odot}$. The dynamical mass in NGC 5595 is $M_{\rm dyn} \sim 1.3 \times 10^{11} M_{\odot}$ as measured from the observed maximum velocity; see Figure 9-Middle panel.

\subsection{NGC 5595: Our VLA B-configuration 20 cm Radio Continuum Emission Map}

    Previously published 20 cm VLA radio continuum observations with an
angular resolution of $\sim 21''$ at FWHM showed an unresolved central source in NGC 5595 elongated into the NE - SW orientation  \citep{con90}, while at an angular resolution of $\sim 7''$ at FWHM the VLA image showed also an unresolved central source with at least three other weaker peaks of emission and weak extended emission surrounding them. All the 20 cm radio continuum emission reported arise from the disk of NGC 5595 \citep{con90}.

    We have produced a new 20 cm radio continuum emission image from our
VLA B-configuration observations with an angular resolution of $\sim 6\farcs11 \times 3\farcs7$ at FWHM (P.A.$\sim -8\degr$ EofN). This continuum image is shown in Figure 13 in contours superimposed on the reproduction of the blue optical continuum 103aO image in grey scale. 
The 20 cm continuum image shows an unresolved dominant central source with a peak flux density of 2.59 mJy at $\alpha \sim 14^h\,24^m\,13\rlap{.}{^s}203$, $\delta \sim -16\degr\,43'\,21\farcs99$. This position coincides with the photometric blue optical continuum 103aO nucleus and the kinematic \hi\ 21 cm center (see Tables 1 and 4). Additionally, there are other 20\,cm continuum sources in what seems a circumnuclear region within an angular distance from the nucleus $R_{\rm N-CNR} \sim 9\farcs5$ or within a linear distance $R_{\rm N-CNR} \sim 1.77$ kpc. There are also 20\,cm continuum sources associated with the north and SW spiral arms within the disk. The total 20\,cm radio continuum flux density in our new image obtained with the VLA in B-configuration is $\sim 81.83$ mJy.

    Our new VLA B-configuration 20\,cm radio continuum image shows
that, at the sensitivity of the observations, all the 20 cm radio continuum emission originates from the optical disk of NGC 5595 (see Figure 13). At the position of the unresolved dominant central 20\,cm radio continuum emission there is no \hi\ 21 cm neutral cold gas emission (see Figures 10). The origin of the 20\,cm radio continuum emission is most likely a mixture of thermal ($T_e \sim 10^4$ K) and optically thin synchrotron processes related to star formation (\hii\ regions) and their evolution (supernovas). There are no 20\,cm radio continuum emission from the upper-most NE and the lower-most SW structures that exhibit \hi\ 21 cm emission as seen in the Moment 0 image (see Figures 10, 11 and 12).

\begin{figure*}[bht]
\begin{center}
\includegraphics[width=11cm,height=11cm]{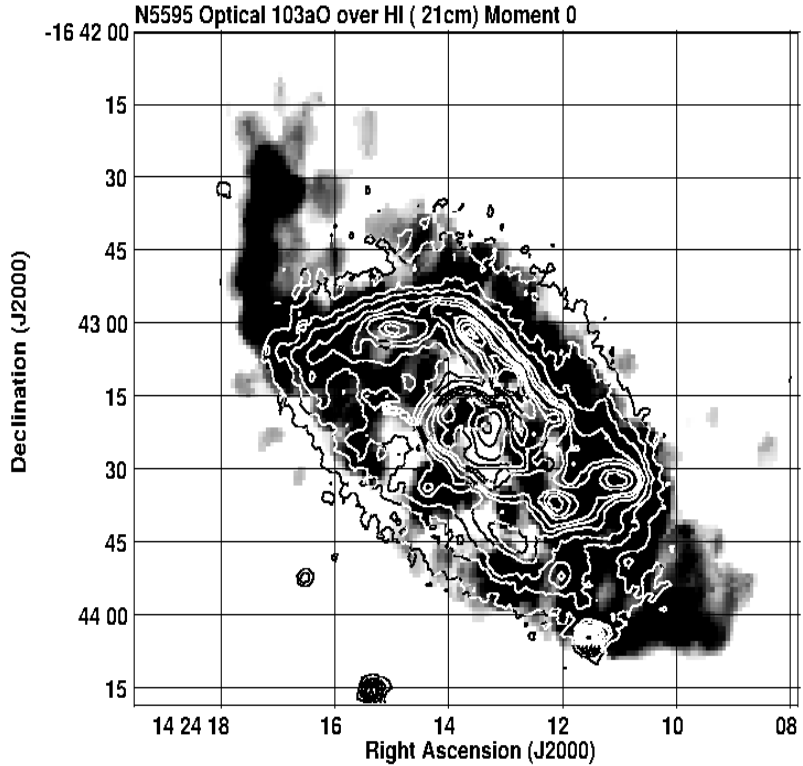}
\figcaption{Reproduction of the 103aO blue optical emission of NGC 5595 \citep{dia09} in contours superimposed on the \hi\ 21 cm Moment 0 image in grey scale. Optical emission is not flux calibrated; its contours are at 4.7,9,13,20,25,30,40,46,53,60,75 and 92 times 1$\sigma$ where 1$\sigma \sim 90$ in arbitrary units proportional to intensity. The grey scale stretch is from 6.6 to 37 mJy beam${^-1}$ km s${^-1}$. Notice that the \hi\ 21 cm NE and SW elongated structures do not have optical continuum emission counterparts.
\label{fig. 12}}
\end{center}
\end{figure*}

\begin{figure*}[bht]
\begin{center}
\includegraphics[width=11cm,height=11cm]{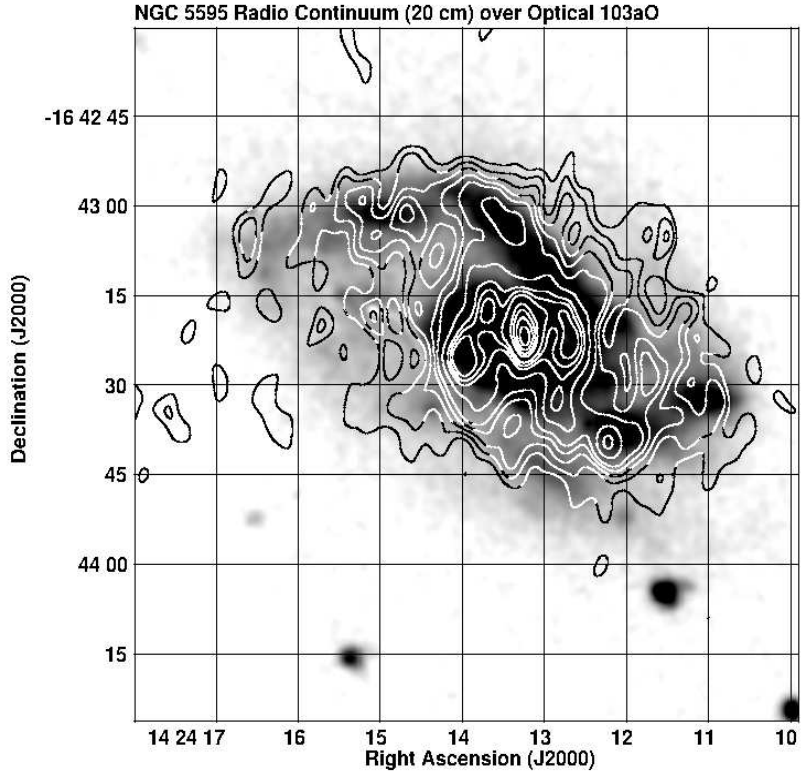}
\figcaption{Our VLA B-configuration 20 cm radio continuum emission from disk galaxy NGC 5595 in contours superimposed on the reproduction of the 103aO blue optical emission from NGC 5595 \citep{dia09} in grey scale. Contours are at 3, 4, 5, 6, 8, 10, 11, 12, 14, 15, 15.6 times 1$\sigma$ where 1$\sigma \sim 167$ $\mu$Jy beam$^{-1}$. The brightest 20 cm radio continuum source coincides with the optical nucleus. Two other sources at a distance $\sim 10''$ from the center toward the east and the west might be part of a circumnuclear structure. There is weak emission from almost all over the optical disk, but no 20 cm radio continuum emission from neither the NE nor the SW \hi\ 21 cm extended structures (see Figure 11).
\label{fig. 13}}
\end{center}
\end{figure*}

\begin{figure*}[bht]
\begin{center}
\includegraphics[width=11cm,height=11cm]{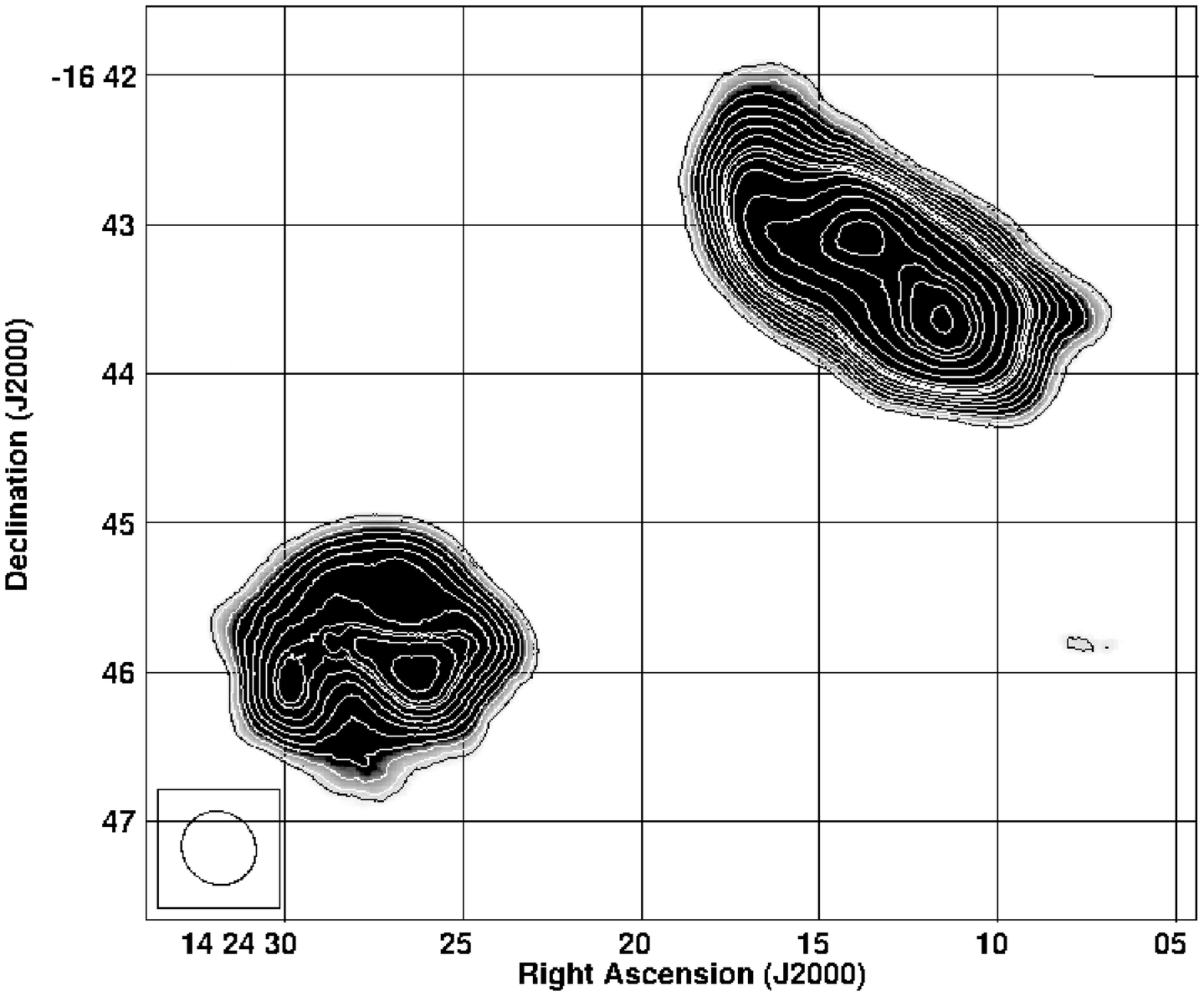}
\figcaption{Our low resolution VLA B-configuration \hi\ 21 cm Moment 0 image in grey scale and contours, obtained using a $u, v$ range restricted between 0 and 5\,k$\lambda$, to search for any \hi\ 21 cm intergalactic extended emission or bridges associated with the disks of NGC 5595 and NGC 5597. NGC 5597 is at south-east, NGC 5595 is at north-west. The grey scale stretch is 250 mJy beam$^{-1}$ km s$^{-1}$ to 450 mJy beam$^{-1}$ km s$^{-1}$. The Moment 0 contours are at 15, 17, 21, 25, 30, 35, 40, 45, 50, 52, 55, 60, 70, 80, 90, 95, 100 and 108 times 1$\sigma$ where 1$\sigma \sim 19.45$ mJy beam$^{-1}$ km s$^{-1}$. No intergalactic \hi\ 21 cm structure or filaments exist between the two galaxies. The synthesized beam at FWHM, which is $\sim 30\farcs7 \times 28\farcs8$ at P.A.$\sim +50\degr$ EofN, is shown at the bottom left corner. }
\label{fig. 14}
\end{center}
\end{figure*}

\section{No Tidal Filamentary Structures in between the Field of the Pair Disk System NGC 5595 and NGC 5597}

    The isolated galaxy pair NGC 5597 - NGC 5595 is in an area of the
universe with very low galaxy volume density \citep{tam85,tul87}, therefore, there is no hot intergalactic gas which could have stripped the atomic gas from outer disk radii. Thus, it seems that both NGC 5595 and NGC 5597 originated with just enough \hi\ atomic neutral gas to form their disks and stars.

\startlongtable
\begin{deluxetable*}{lllccc}
\tabletypesize{\normalsize}
\tablecaption{Kinematical \hi\ 21 cm VLA B-configuration Analysis of the Disk Galaxy Pair NGC 5595 and NGC 5597} \label{tab:table3}
\tablehead{
\colhead{Galaxy} & \colhead{$\alpha$(J2000)} & \colhead{$\delta$(J2000)} & \colhead{Position Angle$^a$} & \colhead{Inclination} & \colhead{$V(\hi)_{{\rm helio}}$} \\
\colhead{name}  & \colhead{$hh~mm~ss.ss$}  & \colhead{$\degr~~' ~~''$} & \colhead {$\degr$ EofN} & \colhead{$\degr$}  & \colhead{km s$^{-1}$}}
\colnumbers
\startdata
NGC 5595 &  $14~24~13.31$ & $-16~43~21.59$ & 237 & 56 & 2702 \\
NGC 5597 &  $14~24~27.16$ & $-16~45~46.64$ & 100 & 36 & 2698 \\
\hline
\enddata
\tablenotetext{a}{Position angle of the semimajor axis of redshifted velocities.}
\end{deluxetable*}

\begin{deluxetable*}{lcc}
\tabletypesize{\normalsize}
\tablecaption{Properties of Disk Galaxy Pair NGC 5595 and NGC 5597} \label{tab:table4}
\tablehead{
\colhead{Data} & \colhead {NGC 5595} & \colhead{NGC 5597} }
\colnumbers
\startdata
Optical diameter & $\sim 1\rlap{.}{'}6$ & $\sim 1\rlap{.}{'}87$\\
m$_{\rm v}$ & 12.6 & 13.1 \\
$V_{\rm sys}$ & 2702 km s$^{-1}$ & 2698 km s$^{-1}$ \\
$M_{\rm HI}$ &  $2.9 \times 10^9~M_{\odot}$ & $1 \times 10^9~M_{\odot}$ \\
$M_{\rm dyn}$ &  $1.3 \times 10^{11}~M_{\odot}$ & $2.6 \times 10^{10}~M_{\odot}$ \\
$L_{\rm FIR}^{\rm IRAS}$ & $2.22 \times 10^{10}~L_{\odot}$ & $2.21 \times 10^{10}~L_{\odot}$ \\ 
P.A.$_{\rm semimajor-axis}^{red}$ & $237\degr$ EofN & $100\degr$ EofN \\
P.A.$_{\rm rotation-axis}$ & $327\degr$ EofN & $190\degr$ EofN \\
Direction$_{\rm rotation-axis}$ & NW & SW \\
Peak flux density (20\,cm) & 2.593 mJy beam$^{-1}$ & 9.02 mJy beam$^{-1}$ \\
Total flux density (20\,cm) & 81.83 mJy & 36.99 mJy \\
\enddata
\end{deluxetable*}

    The blue continuum image (Figure 1) does not show any extended
optical emission or bridges between NGC 5595 and NGC 5597, this fact might indicate that this pair of galaxies are in their early stages of gravitational interaction \citep{too72,mih96,lin22}. 

    However, our VLA B-configuration observations of both disk galaxies
with an angular resolution of $\sim 7\farcs14 \times 4\farcs2$ P.A. $\sim -11\deg$ EofN show that the \hi\ 21 cm emission from NGC 5597 is confined to the extent of its optical disk, while in NGC 5595 the \hi\ 21 cm emission is from its disk, as well as from structures to the NE and the SW that do not have blue optical nor 20 cm continuum counterparts (see Figure 1, Figure 10, and Figure 12). 

    Computer simulations of gas dynamics and starbursts in disk mergers
of similar mass suggest that as the galaxies approach they become severely distorted, forming long tidal tails and a bridge connecting the two disk galaxies \citep{bar90,mih94,mih96,bar98}. A recent retrograde passage dynamical model of the M101 and NGC 5474 reproduces the observations well while suppressing the formation of long tidal tails \citep{lin22}. The spatial distribution of the cold atomic \hi\ 21 cm gas specially towards the NE and SW outside the optical disk in NGC 5595 suggests that the pair system NGC 5595 -- NGC 5597 is in a very early phase of their gravitational interaction with perhaps a retrograde passage \citep{lin22}.

    Table 4 lists the different intrinsic parameters of both galaxies,
including the estimated dynamical masses. The angular separation of the galaxies on the plane of the sky is $3\rlap{.}{'}97$ and the ratio of their dynamical masses is $M_{\rm dyn}^{\rm NGC5595}$/$M_{\rm dyn}^{\rm NGC5597} \sim 5$. 

    From the empirical observational studies of radio continuum emission
from pair of disk galaxies, the central sources in barred galaxies are more powerful than in non-barred ones \citep{hum90}. In our study, the barred galaxy NGC 5597 has a nuclear unresolved 20\,cm radio continuum source that is about 3.5 times more powerful than the similar source in NGC 5595.  Furthermore the innermost central region of NGC 5597 shows H$\alpha$ and 20\,cm radio continuum emissions elongated in the NE - SW orientation (at a P.A. not far from the P.A. of the rotation axis). This innermost region has no \hi\ 21 cm atomic gas emission, while the galaxy shows an \hi\ gas angular velocity curve $\Omega_{\rm gas}$ different than the expected $\Omega_{\rm gas} \propto R^{-1}$ for a spiral disk galaxy, suggesting an excess of mass.  All these observational facts suggest that gas has recently been supplied as fresh fuel into the  barred galaxy NGC 5597, while there are \hi\ 21 cm extended structures to the NE and SW in disk galaxy NGC 5595.

    We have made a low angular resolution Moment 0 image to search for
extended \hi\ 21 cm emission from the field of the pair system NGC 5595 and NGC 5597, using our VLA B-configuration observations, 
restricting the $uv$ range from 0 to only 5k$\lambda$ (Figure 14). This resulted in a synthesized beam with FWHM of $\sim 30\farcs7 \times 28\farcs8$ at a P.A. $\sim +50\degr$ EofN and $1 \sigma \sim 19.45$ mJy beam$^{-1}$ km s$^{-1}$. 

    Figure 14 shows no extended intergalactic \hi\ 21 cm emission from
any filaments or bridges, but only slightly extended emission from the S and SW and W of NGC 5597 and from NE, SE and SW of NGC 5595\footnote{The lack of extended intergalactic \hi\ 21 cm emission from any filaments or bridges may be due to low surface brightness sensitivity. Future VLA \hi\ 21 cm observations with lower angular resolution, deeper integrating, and therefore better surface brightness sensitivity, would clarify the existence of any intergalactic neutral atomic hydrogen gas in the disk galax pair NGC 5595 and NGC 5597.}.

    This negative observational result (no long tidal tails in between
NGC 5595 and NGC 5597) is interesting and important because NGC 5597 and NGC 5595 are not only very close on the plane of the sky, but also physically as they both have very similar recession velocities. Table 4 shows the general properties of both NGC 5595 and NGC 5597. These two galaxies also reside in a low galaxy density environment \citep{tam85,tul87}. The NGC 5597 - NGC 5595 system must be in an early stage of gravitational interaction,
otherwise if it were in an advance stage the NGC 5595 - NGC 5597 pair system would resemble the antennae or the  mice galaxy pairs \citep{too72,bar90,mih96,bar98,lin22}. As a comparison, a long tidal \hi\ 21 cm atomic neutral hydrogen gas structure has been detected in the M51 - NGC 5195 system, which are within 5 arcmin on the plane of the sky \citep{rot90}; extended filamentary \hi\ 21 cm atomic neutral hydrogen gas structures have been detected from the M81 - M82 - NGC 3077 system, which are within 80 arcmin on the plane of the sky \citep{yun94}, and of course the long tidal tails of the antennae system NGC 4038/9 \citep{van79,gor01,hib01}. Computer N-body simulations have fairly reproduced the stellar and \hi\ (21 cm) long tidal tails of the antennae system \citep{bar98}. However, despite much effort no satisfying model is yet available to reproduce the long SE \hi\ 21 cm tidal tail in the M51 system \citep{too78,her90,bar98}.

\section{Summary and Conclusions}

    In this study we have obtained VLA B-configuration \hi\ 21 cm
kinematical data from the close pair disk galaxies NGC 5597 SBc(s), and NGC 5595 Sc(s). We have detected for the first time the existence of cold atomic hydrogen \hi\ 21 cm extended structures (streamers) to the north-east, and south-west of NGC 5595 with no blue, nor red optical continuum, nor 20 cm radio continuum counterparts. They may be filamentary features as a result of recent gravitational tidal interaction with its neighbor NGC 5597 to the south-east. 

    We were able to get the best fit velocity fields, rotation curves
($V$ vs. $R$), and angular velocities ($\Omega_{\rm gas}$ vs. $R$) for both disk galaxies NGC 5595 and NGC 5597 from their heliocentric \hi\ 21 cm recession velocities, assuming gas is in circular orbits around the nucleus. We have presented the \hi\ 21 cm spectra for NGC 5597 and NGC 5595, estimated the \hi\ 21 cm gas mass from many clouds in both NGC 5595 and NGC 5597, as well as the total \hi\ 21 cm gas mass in both galaxies. 

    Our \hi\ 21 cm fitted parameters indicate that the disk rotation
axis in NGC 5597 projected on the plane of the sky is at P.A.$\sim 190$ EofN pointing into the SW direction, thus the hemisphere from NW clockwise towards SE is closer to the observer, while the disk rotation axis in NGC 5595 projected on the plane of the sky is at P.A.$\sim 327$ EofN pointing into the NW direction, thus the NE hemisphere clockwise towards SW is closer to the observer.

    We have also presented new 20 cm radio continuum emission images
from both NGC 5595 and NGC 5597. In particular in NGC 5597, our image shows an unresolved central emission peak with an elongated structure that coincides with previously published H$\alpha$+N[II] line emission. Both emissions come from the innermost central region where there is no cold hydrogen atomic gas in NGC 5597. The existence of H$\alpha$ and 20cm radio continuum, and the lack of \hi\ 21 cm at the center of NGC 5597 are very similar to the distribution observed in other nearby barred galaxies and might suggest a central SMBH with mass of few times 10$^6 M_{\odot}$.

    Finally, we have made low resolution ($\approx 30''$ at FWHM) \hi\
21 cm image of the field of the pair disk galaxy system NGC 5595 and NGC 5597 and did not detect any extended intergalactic \hi\ 21 cm tails or bridges in between the disk galaxies.

\vspace{5mm}
\facilities{EVLA}

\software{
AIPS: \citep{gre03},
CASA \citep{mcm07}
}

\end{document}